\documentclass[12pt]{article}
\pdfoutput=1
\usepackage{graphicx}
\usepackage{epstopdf}
\usepackage{amsmath}
\usepackage{amsfonts}
\usepackage{amssymb}
\usepackage{color}
\usepackage{mathrsfs}
\usepackage[utf8]{inputenc}
\usepackage{subcaption}

\usepackage[font={footnotesize}]{caption}

\pdfoutput=1

\usepackage[english]{babel}

\usepackage[colorlinks,bookmarks]{hyperref}
\definecolor{linkblue}{rgb}{0,0,0.8}
\definecolor{linkgreen}{rgb}{0,0.5,0}

\hypersetup{pdfpagemode=UseNone, pdfstartview=FitH, linkcolor=linkblue, %
            citecolor=linkgreen, urlcolor=linkblue}

\newcommand{\Msun}{M_{\odot}}

\newcommand{\website}{\url{http://web.stanford.edu/~senatore/}}

\numberwithin{equation}{section}

\newcommand{\hinvMpc}{\,h\, {\rm Mpc}^{-1}\,}

\newcommand{\bea}{\begin{eqnarray}}
\newcommand{\eea}{\end{eqnarray}}
\newcommand{\be}{\begin{equation}}
\newcommand{\ee}{\end{equation}}
\newcommand{\fr}[2]{\frac{ #1}{#2}}
\newcommand{\n}{\nonumber \\}

\newcommand{\kv}{\vec{k}}
\newcommand{\qv}{\vec{q}}

\newcommand{\knl}{k_{\rm NL}}

\newcommand{\eqn}[1]{Eq.~(\ref{#1})}

\newcommand{\unitsk}{ h \ { \rm Mpc^{-1}}}
\newcommand{\unitskinv}{ h^{-1} { \rm Mpc}}
\newcommand{\maxx}{\rm max}

\newcommand{\delhr}{\delta_{h,r}}
\newcommand{\delh}{\delta_{h}}

\newcommand{\q}{\vec{q}}

\newcommand{\dzdsq}{\fr{\partial_z}{\partial^2}}
\newcommand{\kz}{k_{z}}
\newcommand{\km}{k_{\rm M}}
\newcommand{\xv}{\vec{x}}
\newcommand{\xf}{\vec{x}_{\rm fl}}
\newcommand{\sym}{{\rm sym}}

\newcommand{\pd}{\partial}

\newcommand{\Comment}[1]{{}}

\begin{document}
\def\thefootnote{\fnsymbol{footnote}}

\setcounter{page}{1} \baselineskip=15.5pt \thispagestyle{empty}

\begin{flushright}
\end{flushright}

\begin{center}

{\Large \bf  Biased Tracers in Redshift Space \\[0.3cm] in the EFT of Large-Scale Structure  \\[0.7cm]}
{\large  Ashley Perko$^{1,2}$, Leonardo Senatore$^{1,2, 3}$,\\[0.3cm] Elise Jennings$^{4,5}$, and Risa H. Wechsler$^{2,3}$}
\\[0.7cm]

{\normalsize { \sl $^{1}$ Stanford Institute for Theoretical Physics,\\ Stanford University, Stanford, CA 94306}}\\
\vspace{.3cm}

{\normalsize { \sl $^{2}$ Department of Physics, \\
Stanford University, Stanford, CA 94305}}\\
\vspace{.3cm}

{\normalsize { \sl $^{3}$ Kavli Institute for Particle Astrophysics and Cosmology and Dept. of Particle Physics and Astrophysics, SLAC, Menlo Park, CA 94025}}\\
\vspace{.3cm}

{\normalsize { \sl $^{4}$ Center for Particle Astrophysics, Fermi National Accelerator Laboratory MS209, \\
 P.O. Box 500, Kirk Rd. \& Pine St., Batavia, IL 60510-0500}}\\
\vspace{.3cm}

{\normalsize { \sl $^{5}$ Kavli Institute for Cosmological Physics, \\
Enrico Fermi Institute, University of Chicago, Chicago, IL 60637}}\\
\vspace{.3cm}

\end{center}

\vspace{.8cm}

%
%
%

\hrule \vspace{0.3cm}
{\small  \noindent \textbf{Abstract} \\[0.3cm]
\noindent  The Effective Field Theory of Large-Scale Structure (EFTofLSS) provides a novel formalism that is able to accurately predict the clustering of large-scale structure (LSS) in the mildly non-linear regime. Here we provide the first computation of the power spectrum of biased tracers in redshift space at one loop order, and we make the associated code publicly available. We compare the multipoles $\ell=0,2$ of the redshift-space halo power spectrum, together with the real-space matter and halo power spectra, with data from numerical simulations at $z=0.67$. For the samples we compare to, which have a number density of $\bar n=3.8 \cdot 10^{-2}(\hinvMpc)^3$ and $\bar n=3.9 \cdot 10^{-4}(\hinvMpc)^3$, we find that the calculation at one-loop order matches numerical measurements to within a few percent up to $k\simeq 0.43\hinvMpc$, a significant improvement with respect to former techniques. By performing the so-called IR-resummation, we find that the Baryon Acoustic Oscillation peak is accurately reproduced. Based on the results presented here, long-wavelength statistics that are routinely observed in LSS surveys can be finally computed in the EFTofLSS. This formalism thus is ready to start to be compared directly to observational data. 
 \vspace{0.3cm}
\hrule

\def\thefootnote{\arabic{footnote}}
\setcounter{footnote}{0}

%
%
%
%
\newpage
\tableofcontents
\section{Introduction}

In the next decade, large-scale structure data, collected through galaxy surveys, the CMB and possibly 21cm emission, will be essential for our progress in understanding the first instants of the universe and the late time growth of structures. In order to fully profit from this collection of data, it will be important to be able to make theoretical predictions with comparable or better accuracy than the errors in the observations. Current techniques have not yet achieved this goal in many regimes. Numerical simulations have been the leading tool for predicting the clustering of large-scale structure so far. However, keeping pace with the data to achieve the required accuracy over the full range of parameter space needed will be challenging~\cite{Schneider:2015yka}.  On the analytic side, in recent years the Effective Field Theory of Large-Scale Structure~\footnote{The Effective Field Theory of Large-Scale Structure is the same as the Effective Field Theory of Large Scale Structures. The only difference is that, as we are now moving closer to observations, we are thought to use a more standard terminology as well as to use more observational-related information, such as the details of the samples that we give already in the abstract. } (EFTofLSS)~\cite{Baumann:2010tm,Carrasco:2012cv,Porto:2013qua,Senatore:2014via} has provided a tool where predictions in the mildly non-linear regime can be delivered with an at-least-in-principle arbitrary theoretical precision, and most probably with a precision that should be enough for next generation experiments~\cite{Baumann:2010tm,Carrasco:2012cv,Porto:2013qua,Senatore:2014via,Carrasco:2013sva,Carrasco:2013mua,Pajer:2013jj,Carroll:2013oxa,Mercolli:2013bsa,Angulo:2014tfa,Baldauf:2014qfa,Senatore:2014eva,Senatore:2014vja,Lewandowski:2014rca,Mirbabayi:2014zca,Foreman:2015uva,Angulo:2015eqa,McQuinn:2015tva,Assassi:2015jqa,Baldauf:2015tla,Baldauf:2015xfa,Foreman:2015lca,Baldauf:2015aha,Baldauf:2015zga,Bertolini:2015fya,Bertolini:2016bmt,Assassi:2015fma,Lewandowski:2015ziq,Cataneo:2016suz,Bertolini:2016hxg}.

This paper represents the culmination of a journey: we bring the theoretical construction of the EFTofLSS all the way to computing statistics of the galaxy distribution in redshift space, which can be compared directly with observational data from large-scale structure surveys. Computing such observables is indeed a complex endeavor. In order to predict a given correlation function of galaxies, one needs first to predict correlation functions of dark matter, then of biased tracers, and then one has to map these correlation functions into redshift space. Each of these steps requires novel calculations as well as understanding how the predictions can be made insensitive to the uncontrolled short distance fluctuations that affects even long distance fluctuations, through the so-called process of renormalization. Furthermore, one needs to understand and implement the procedure that is called IR-resummation~\cite{Senatore:2014via}, and that amounts to non-linearly solving for the effect of the long wavelength displacement modes, which are crucial to correctly predicting the BAO peak. 

In the few years since the introduction of the EFTofLSS, each of the steps necessary to connect the EFTofLSS to observations has been tackled one by one. Several studies have been performed on dark matter correlation functions.  For example, the EFTofLSS community has studied the dark matter density two-point function~\cite{Carrasco:2012cv,Senatore:2014via,Carrasco:2013mua,Foreman:2015lca,Baldauf:2015aha}, three-point function~\cite{Angulo:2014tfa,Baldauf:2014qfa}, four-point function (which includes the covariance of the power spectrum)~\cite{Bertolini:2015fya,Bertolini:2016bmt}, the dark matter momentum power spectrum~\cite{Senatore:2014via,Baldauf:2015aha},  the displacement field~\cite{Baldauf:2014qfa}, and the vorticity slope~\cite{Carrasco:2013mua,Hahn:2014lca}. The effects of baryons on the matter correlation functions have been incorporated in the EFTofLSS in~\cite{Lewandowski:2014rca}.  Moving to biased tracers, the extension of the EFTofLSS to describe these objects has been carried out in~\cite{Senatore:2014vja}, and the predictions compared to data for the power spectrum and bispectrum (including all mixed correlation functions between matter and halos) in~\cite{Angulo:2015eqa,Fujita:2016dne}. Redshift-space distortions have been included in the EFTofLSS in~\cite{Senatore:2014vja}, and they have been compared to numerical data for matter power spectra in~\cite{Lewandowski:2015ziq}. Methods to measure the parameters of the EFTofLSS from small numerical simulations have been developed in~\cite{Carrasco:2012cv,Baldauf:2011bh,Baldauf:2015vio,Lazeyras:2015lgp,Bertolini:2016hxg}. The IR-resummation was implemented and compared to numerical data in~\cite{Senatore:2014via} for dark matter, then extended to halos in~\cite{Senatore:2014vja} and compared to halo data in~\cite{Angulo:2015eqa}, and then extended to dark matter in redshift space and compared to simulated datasets in~\cite{Senatore:2014vja,Lewandowski:2015ziq}. The impact of primordial non-Gaussianity on large-scale structure observables~\cite{Angulo:2015eqa,Assassi:2015jqa,Assassi:2015fma,Lewandowski:2015ziq} has also been recently included. Fast implementations of the predictions of the EFTofLSS, which allows us to efficiently explore their dependence on various cosmological parameters, have been recently developed in~\cite{Cataneo:2016suz}, with public codes available at the following website~\footnote{\label{website}\website}. 

After the completion of this collection of works, a final step is needed to complete the development of the theory so that it is ready to be compared with observations of large-scale structure surveys: to compute the power spectrum (or the correlation function, which is just its Fourier transform) of dark matter halos in redshift space. This calculation is the primary focus of the present work. We compute the multipoles $\ell=0$ and $2$ of the power spectrum for halos in redshift space at one loop, perform the IR-resummation, and, together with the real-space dark matter and halo power spectra, we compare to their measurements in numerical simulations.~\footnote{We stress that the theory at the order at which we work predicts, without any additional parameters, other observables, such as for example the $\ell=4,6$ multipoles or the tree-level bispectra. We leave a detailed examination of higher multipoles and of bispectra for a galaxy sample size expected in future surveys to future work, when the measurement of these multipoles will be available to us. } The codes that allow us to obtain the results we present in this paper are made publicly available at the website given in footnote~\ref{website}. Here we use measurements of the power spectra from one of the {\em DarkSky} N-body simulations~\cite{Skillman:2014qca, Jennings:2015lea} at redshift $z\simeq 0.67$, and find that the one-loop predictions of the EFTofLSS match the numerical data up to about $k\simeq 0.43\hinvMpc$. Reaching these scales is very encouraging for the future prospects of extracting cosmological information from galaxy surveys.

If this work, in a sense, represents the completion of a journey for the development of the theory of the EFTofLSS, it also represents the beginning of a new, and more important, journey. Now we are finally ready to start to apply the EFTofLSS to the cosmological measurements that are accessible from large-scale structure surveys. This will enable us to ascertain how much novel information about the universe will be available to us by interpreting these observables through the framework of the EFTofLSS. Of course, as the one we just completed, this novel journey will have its own challenges.  For example, in order to analyze data through the predictions of the EFTofLSS, we will need to understand how to systematically take into account and measure the free parameters that are present in the EFTofLSS, as well as how to account for new systematic errors that may be introduced by the theoretical errors in the calculations, as well as how different galaxy populations affect the results. We are sure that some of these challenges will force us to go back on the theory side, and understand better the theory of the EFTofLSS, so the two journeys, the theoretical one and the observational one, will not  be completely detached. Regardless, the novel and very exciting phase of applying the EFTofLSS to the direct interpretation of observational data can finally begin.

%
%
%
%
\section{Biased tracers in real space}
\subsection{Review of $\delta_h$ perturbative expansion}
The effective field theory of biased tracers of large-scale structure in real space was previously studied in \cite{Senatore:2014eva,Angulo:2015eqa,Fujita:2016dne}. We will review that treatment here, emphasizing the elements needed for the computation in redshift space. Since dark matter halos and galaxies form due to gravitational collapse, they are biased tracers of the underlying dark matter fields. This means that the density and velocity of halos depend on the dark matter density $\delta$ and velocity $v^i$, as well as the second derivative of the gravitational potential $\partial^2\phi$. Since we are only interested in describing correlation functions on scales much larger than a single halo, we can write the halo fields as an expansion in  spatial derivatives of $\delta$, $v^i$, and $\partial^2\phi$, suppressed by the scale of the halos, $\km^{-1}$. However, a similar expansion cannot be made for the time dependence of the halos. This is due to the fact that the typical formation time of the halos is not parametrically larger than the timescales of interest, which are of order $H^{-1}$. Thus, in effective field theory the halo density is written as an expansion in spatial derivatives of $\delta$, $v^i$, and $\partial^2\phi$, each of which are integrated over the formation history of the halo, with a time-dependent kernel describing the short-wavelength dynamics of halo formation. For example, the halo density is written as
\bea
\delta_h(\xv,t)&=& \int^t dt' \ H(t') \left[ c_{\pd^2\phi}(t,t') \frac{\pd^2 \phi(\xv_{\rm fl},t)}{H('t)^2}+ c_{\pd_i v^i}(t,t')\frac{\pd_i v^i(\xv_{\rm fl},t')}{H(t')} \right. \n
&&  \left.  + c_{\pd_i \pd_j \phi \pd^i \pd^j \phi}(t,t')\frac{\pd_i \pd_j \phi \pd^i \pd^j \phi(\xv_{\rm fl}, t')}{H(t')^4} + c_{\partial^2 \delta}(t,t') \frac{\partial^2}{\km^2} \delta(\xf, t')+ \  \ldots \ \right] \ ,
\eea
where the $c_i(t,t')$ are the undetermined kernels that encode the UV physics of halo formation and $\ldots$ includes terms that are higher order in perturbations, as well as higher derivative terms, which are suppressed by $\pd^2_{x_{\rm fl}}/\km^2$, and stochastic terms, which we will discuss in detail in Section \ref{stochhalo}. The fields in the expansion depend on the fluid position $\xv_{\rm fl}$, which is recursively defined as 
\be
\xv_{\rm fl}(\xv,\tau,\tau')=\xv-\int_{\tau'}^\tau d \tau'' \vec{v}(\tau,\tau'') \ ,
\ee
with $\tau$ the conformal time. They depend on this coordinate rather than $\xv$ because $\xv_{\rm fl}$ is the parameter that follows the matter forming the halo.

The halo velocity $v_h^i$ can be expanded in a similar way, but since it does not transform like a scalar, we have to be careful about the transformation properties of the fields in the expansion. Consider transforming to the inertial frame of a long wavelength mode of the dark matter, where $v^i_m=0$. In that frame, the dark matter is stationary by construction, which means that that the halos are also stationary. Going back to the original frame, this means that diffeomorphism invariance requires that the halo velocity must be equal to the dark matter velocity at linear order. Diffeomorphism invariance also implies that gravitational potential must appear with at least two derivatives. This is because $\phi$ and $\pd_i \phi$ can be transformed away by a change of coordinates, so $\pd^2 \phi$ is the first physically relevant field in the derivative expansion of $\phi$. This means that when we write the terms depending on $\delta$ and $\pd^2\phi$ in the expansion for $v^i_h$, each has at least one factor of $\pd^i_{x_{\rm fl}}/\km$ suppressing it in order to have the correct transformation properties. Thus at leading order in derivatives and neglecting the stochastic terms for now, the expansion for $v^i_h$ in terms of dark matter fields is simply
\be
v^i_h(\xv,t)=v^i(\xv_{\rm fl},t) + \int^t dt' c_{\partial^i \delta}(t,t')\frac{ \partial^i}{\km} \delta(\xf, t')+ \ \ldots  \ .
\label{vel}
\ee
Let us focus for the moment on the dark matter fields. Instead of considering $v^i$ directly, it will be more convenient to consider the velocity divergence, defined as $\theta \equiv -\frac{D}{D'}\pd_i v^i$, and the velocity vorticity $\pd_i v_j$. Notice that with this normalization of $\theta$, at linear level $\theta^{(1)}=\delta^{(1)}$. Taking advantage of this, we will define the new variable $\eta=\theta-\delta$, which is nonzero only starting at second order.  From the equations of motion we find that $\eta^{(2)}=\frac{2}{7}(s^{(1)})^2-\frac{4}{21}(\delta^{(1)})^2$ \cite{McDonald:2009dh,Senatore:2014eva,Angulo:2015eqa}, so we can define a parameter $\psi$ that is nonzero starting only at third order, 
 \be
 \psi \equiv \eta-\frac{2}{7}s^2+\frac{4}{21} \delta^2 \ .
  \ee
Next we will define the traceless tidal tensor $s_{ij}\equiv \pd_i \pd_j \phi -\frac{1}{3}\delta_{ij} \delta$, where the potential $\phi$ is defined via the Poisson equation $\pd^2 \phi=\delta$. We will also define a new field $t_{ij}$ in place of $\partial_i v_j$, which is given by
\be
t_{ij} \equiv -\frac{D}{D'} \pd_i v_j-\frac{1}{3}\delta_{ij}\theta-s_{ij} \ .
\ee
With this definition, $t_{ij}$ is nonzero starting at second order, and can be considered symmetric at the approximation to which we are working because vorticity is not generated until very high orders in perturbation theory \cite{Carrasco:2013mua,Mercolli:2013bsa}.

In summary, instead of $\delta$, $v^i$, and $\partial^2\phi$, our dynamical variables in perturbation theory are $\delta$, $s_{ij}$, $t_{ij}$,  and $\psi$. Now we can find the halo density in terms of these fields by forming all possible combinations of $\delta$, $s_{ij}$, $t_{ij}$, and $\psi$ that are rotationally-invariant, and which are integrated over kernels with support over the last Hubble time. In these new variables, the expansion for the halo density to third order in perturbation theory is
\bea
\delta_h(\xv,t)&=&\int^t dt' H(t') \Bigl(  c_\delta(t,t') \delta(\xv_{\rm fl},t')+  c_\delta^2(t,t') \delta(\xv_{\rm fl},t')^2+  c_s^2(t,t') s_{ij}(\xv_{\rm fl},t')s^{ij}(\xv_{\rm fl},t') \n 
&&  +c_\delta^3(t,t') \delta(\xv_{\rm fl},t')^3  +c_{\delta s^2}(t,t') \delta(\xv_{\rm fl},t')s_{ij}(\xv_{\rm fl},t')s^{ij}(\xv_{\rm fl},t')+ c_\psi(t,t') \psi(\xv_{\rm fl},t') \n
&& + c_{st}(t,t') s_{ij}(\xv_{\rm fl},t')t^{ij}(\xv_{\rm fl},t') +c_{\delta s^3}(t,t') \delta(\xv_{\rm fl},t')s_{ij}(\xv_{\rm fl},t')s^{i}_k(\xv_{\rm fl},t')s^{jk}(\xv_{\rm fl},t') \n
&& +  c_{\partial^2 \delta}(t,t') \frac{\partial^2_{\xf} }{\km^2}\delta(\xf,t')+ \ \ldots \  \Bigr) \ ,
\label{convo}
\eea
where again $\ldots$ includes stochastic and higher-derivative terms \cite{Senatore:2014eva}. 

The integrals in time in \eqn{convo}, which contain the time-dependent kernels and the growth factor, can be done symbolically to give new, ``effectively local", time-dependent coefficients. When we do this symbolic integral and go to Fourier space,  \eqn{convo} becomes
\bea
\delta_{A} &\equiv& c^{(A)}_{\delta,1}\delta^{(1)}+ c^{(A)}_{\delta,2}\delta^{(2)}+ c^{(A)}_{\delta,3}\delta^{(3)}+(c^{(A)}_{\delta,1}-c^{(A)}_{\delta,2})[\partial_i \delta^{(1)}\frac{\partial_i}{\partial^2}\theta^{(1)}]  \n
&& +(c^{(A)}_{\delta,2}-c^{(A)}_{\delta,3})[\partial_i \delta^{(2)}\frac{\partial_i}{\partial^2}\theta^{(1)}]  +\frac{1}{2}(c^{(A)}_{\delta,1}-c^{(A)}_{\delta,3})[\partial_i \delta^{(1)} \frac{\partial_i}{\partial^2}\theta^{(2)}] \n
&& +\left(\frac{1}{2}(c^{(A)}_{\delta,1}+c^{(A)}_{\delta,3})-c^{(A)}_{\delta,2} \right) \left([\partial_i \delta^{(1)}\frac{\partial^i\partial_j}{\partial^2}\theta^{(1)}\frac{\partial_j}{\partial^2}\theta^{(1)}]+[\partial_{i}\partial_j \delta^{(1)}\frac{\partial_i}{\partial^2}\theta^{(1)}\frac{\partial_i}{\partial^2}\theta^{(1)}]\right)   +c^{(A)}_{\delta^2,1}[\delta^2]^{(2)}\n
&&+c^{(A)}_{\delta^2,2}[\delta^2]^{(3)}-2(c^{(A)}_{\delta^2,1}-c^{(A)}_{\delta^2,2})
[\delta^{(1)}\partial_i \delta^{(1)}\frac{\partial_i}{\partial^2}\theta^{(1)}]+ c^{(A)}_{\delta^3}[\delta^3]^{(3)}
+c^{(A)}_{s^2,1}[s^2]^{(2)}+c^{(A)}_{s^2,2}[s^2]^{(3)}  \n
&&-2(c^{(A)}_{s^2,1}-c^{(A)}_{s^2,2})[s_{lm}^{(1)}\partial_i s^{lm, (1)}\frac{\partial_i}{\partial^2}\theta^{(1)}]+c^{(A)}_{st}[st]^{(3)}+c^{(A)}_{\psi}\psi^{(3)}+c^{(A)}_{\delta s^2}[\delta s^2]^{(3)}+c^{(A)}_{s^3}[s^3]^{(3)} \n
&&  + \ \ldots \ , 
\label{haloexp}
\eea
where the terms in brackets involving spatial derivatives of $\delta$ and $\theta$ arise from Taylor expanding $\xv_{\rm fl}$ around $\xv$ up to third order, and we have left off the stochastic terms and counter-terms for now. The superscript $A$ refers to the specific halo population, because the coefficients will be different for different halo (or galaxy) populations.

To solve for $\delta_h$, we will expand the dark matter fields in perturbations. The higher order fields for the dark matter are given in terms of the linear fields by integrals in momenta with the standard SPT kernels $F^{(n)}$ and $G^{(n)}$, defined as
\bea
\delta^{(n)}(\kv) &=& \int d^3 q_1 \ldots d^3 q_n  \ F^{(n)}(\q_1, \ldots , \q_n)  \delta^3_D(\kv-\qv_1 \ldots - \qv_n) \delta^{(1)}(\qv_1) \ldots \delta^{(1)}(\qv_n) \n
\theta^{(n)}(\kv) &=& \int d^3 q_1 \ldots d^3 q_n \ G^{(n)}(\q_1, \ldots , \q_n) \delta^3_D(\kv-\qv_1 \ldots - \qv_n) \delta^{(1)}(\qv_1) \ldots \delta^{(1)}(\qv_n) \ ,
\label{dmexp}
\eea
plus counter-terms and stochastic terms. Using the expansions in \eqn{dmexp}, we can express each term in \eqn{haloexp} as an integral over factors of $\delta^{(1)}$, i.e. the linear dark matter field, with the generalized halo kernels defined as
\bea
\delta^{(n)}_{A}(\kv) &=& \int d^3 q_1 \ldots d^3 q_n  K^{(n)}_{A}(\q_1, \ldots , \q_n) _{\sym} \delta^3_D(\kv-\qv_1 \ldots - \qv_n) \delta^{(1)}(\qv_1) \ldots \delta^{(1)}(\qv_n)  \ .
\label{hkernel}
\eea
The full halo field up to third order in perturbation theory can now be written as:
\be
\delta_{A}=\delta_{A}^{(1)}+\delta_{A}^{(2)}+\delta_{A}^{(3)} + \delta_{A}^{(3,{\rm ct})}+ \delta_{A}^{(\epsilon)} \ ,
\label{exppert}
\ee
where $\delta_{A}^{(1)}$, $\delta_{A}^{(2)}$, and $\delta_{A}^{(3)}$ are given by the kernels in \eqn{hkernel}, $\delta_{A}^{(\epsilon)}$ represents the halo stochastic terms that we will discuss later in Section \ref{stochhalo}, and $\delta_{A}^{(3,{\rm ct})}=c_{\rm ct}^{(A)}\delta^{(3,{\rm ct})}$ is the biased dark matter density counter-term, which includes a contribution both from $\delta^{(3,{\rm ct})}$, the dark matter counter-term, and from the higher-derivative bias $\partial^2_{\xf} \delta$, because it is degenerate with $\delta^{(3,{\rm ct})}$.

The explicit expressions for the $K^{(n)}_{A}$ are given in \cite{Angulo:2015eqa}. In \eqn{haloexp} it appears that there are twelve bias coefficients that must be fit to observations $\left( c^{(A)}_{\delta,1} \right.$, $c^{(A)}_{\delta,2}$, $c^{(A)}_{\delta,3}$, $c^{(A)}_{\delta^2,1}$, $c^{(A)}_{\delta^2,2}$, $c^{(A)}_{\delta^3}$ $c^{(A)}_{s^2,1}$, $c^{(A)}_{s^2,2}$, $c^{(A)}_{st}$, $c^{(A)}_{\psi}$, $c^{(A)}_{\delta s^2}$, and $\left.c^{(A)}_{s^3} \right)$. However, the operators multiplying these coefficients, which were computed in \cite{Angulo:2015eqa} and are given explicitly in \eqn{app:second} and \eqn{app:third} of Appendix \ref{appendixa}, are not linearly independent, so in fact this is an over-counting, and there are really eight independent bias parameters. There are yet more degeneracies that appear at the level of the power spectrum, and in the end we will have just four bias parameters for the power spectrum at one loop.  This is an accidental cancellation, which does not occur generically in all observables or for higher loops. The details of the degeneracy of parameters that occurs at one loop in the halo power spectrum are given in Appendix \ref{sec3}.

%
%
%
%

\subsection{The velocity divergence as a biased density tracer}
The halo kernels discussed in the previous section were derived in \cite{Angulo:2015eqa} in order to calculate the power spectrum of halos in real space. There the expansion for $\theta_h$ was not needed because correlation functions of $\theta_h$ were not computed. However, in order to compute the power spectrum of $\delta_h$ in redshift space, we will need the correlations of $\theta_h$ because the transformation to redshift space involves the velocity. Thus we need to compute the analogous kernels for $\theta_h$.

We know from \eqn{vel} that due to diffeomorphism invariance, the expansion for the halo velocity divergence is simply
\be
\theta_h(\xv,t)=\theta(\xv,t)+ \int^t dt' \bar{c}_{ \partial^2 \delta }(t,t')\frac{ \partial^2_{\xf}}{\km^2}  \delta(\xf,t) + \ldots \ ,
\label{velexp}
\ee
neglecting the stochastic terms which we will comment on in the next section. Expanding in perturbations up to third order, $\theta =\theta^{(1)}+ \theta^{(2)}+ \theta^{(3)}$, and using the linear equations of motion and the parameters defined in the previous section, we find
\bea
 \theta^{(1)} &=& \delta^{(1)} \n
 \theta^{(2)} &\equiv& \delta^{(2)}+ \eta^{(2)}=\delta^{(2)}+\frac{2}{7} (s^2)^{(2)}-\frac{4}{21}(\delta^2)^{(2)} \n
 \theta^{(3)} &\equiv& \delta^{(3)}+ \eta^{(3)}= \delta^{(3)}+\psi^{(3)}+\frac{2}{7}(s^2)^{(3)}+\frac{4}{21}(\delta^2)^{(3)} \ ,
\label{tthetaeq}
\eea
which means that the expansion for $\theta_h$ can be written as:
\bea
\theta_{h} &\equiv& \delta^{(1)}+ \delta^{(2)}+ \delta^{(3)} -\frac{4}{21}[\delta^2]^{(2)}-\frac{4}{21}[\delta^2]^{(3)}+\frac{2}{7}[s^2]^{(2)}+\frac{2}{7}[s^2]^{(3)} +\psi^{(3)} \n
&& + \theta_h^{(3,{\rm ct})}+ \ldots \ ,
\label{general}
\eea
where we have neglected stochastic terms and $\theta_h^{(3,{\rm ct})}$ again contains the counter-term from dark matter as well as a contribution from the higher-derivative term  $\partial^2_{\xf}\delta$ in \eqn{velexp}. Notice that \eqn{general} takes the same form as the expression for $\delta_A$ in \eqn{haloexp}, but with the following  specific values for the coefficients: 
\bea
&&c^{(A=\theta_h)}_{\delta_1} =c^{(A=\theta_h)}_{\delta_2}=c^{(A=\theta_h)}_{\delta_3}=c^{(A=\theta_h)}_{\psi}=1 \n
&&c^{(A=\theta_h)}_{s^2,1} =c^{(A=\theta_h)}_{s^2,2}=\frac{2}{7} \n
&&c^{(A=\theta_h)}_{\delta^2,1}=c^{(A=\theta_h)}_{\delta^2,2}=-\frac{4}{21} \n
&&c^{(A=\theta_h)}_{st} =c^{(A=\theta_h)}_{\delta^3}=c^{(A=\theta_h)}_{\delta s^2}=c^{(A=\theta_h)}_{s^3}=0 \ .
\label{coefs}
\eea
This is non trivial, and it happens because the evolution of the dark matter is local, given that at tree level the speed of sound vanishes. Therefore, since the expansion for the halo density already contained all possible {\it spatially-local} terms consistent with the symmetries, the expression for the velocity is simply a special case of that expansion. In essence, this is the same reason why we could use a spatially-local expansion for halos~\cite{Senatore:2014vja}. There are no free bias coefficients in the expression for $\theta_h$ except for the counter-term parameter because of the lack of a linear bias in \eqn{velexp}. Therefore, for the purposes of this calculation, we can think of the velocity divergence field as a special species of halo with fixed coefficients, which we will denote as $\delta_{A}$ with $A=\theta_h$. Now instead of a separate expansion for $\theta_h$, we can simply use the expansion for halos in \eqn{exppert} but with the coefficients given in \eqn{coefs}.

\subsection{Stochastic halo bias}\label{stochhalo}

So far we have neglected the contribution of stochastic bias. Since the effective theory is defined by smoothing over the modes with wavelength shorter than a given cutoff $\Lambda^{-1}$, in general there are stochastic terms due to the fact that there is difference between a given realization of the long wavelength mode in the smoothed region and its expectation value. The resulting stochastic field $\epsilon(\xv,t)$ is expected to be Poisson distributed, to have zero mean and to correlate only with itself and not the other perturbative fields \cite{Carrasco:2012cv,Carrasco:2013mua}. In the case of dark matter, mass and momentum conservation forces the stochastic term to come into the stress tensor with two derivatives, $\Delta \tau_{stoch}^{ij} \sim  \partial^i \partial^j \epsilon(x,t)$, so in the power spectrum the stochastic term is suppressed by $(k/k_{\rm NL})^4$ \cite{Carrasco:2012cv,Carrasco:2013mua}. However, this is no longer the case for halos because their mass and momentum is not conserved due to halo mergers. Thus there will be a stochastic contribution at order $k^0$, which by dimensional analysis scales like $ \langle \epsilon \epsilon \rangle_k \sim (2\pi/{k_0})^3\sim 1/\bar n $, where $k_0$ is the inverse of the typical halo spacing and $\bar n$ is therefore the typical halo density. As discussed in \cite{Baldauf:2015zga}, its typical size can be roughly estimated as
\be
\langle \epsilon \epsilon \rangle_k \sim \frac{1}{\bar{n}_W} =\int dM \ \frac{dn}{dM}\frac{M^2}{\rho_b^2} \ ,
\label{meansq}
\ee 
where $M$ is the mass of the halo, $\rho_b$ is the background matter density, and $dn/dM$ is the halo mass function.

Stochastic terms appearing in the expansion for $\delta_h$ include:
\bea
\delta_h^{(\epsilon)}&=&\left(d_1 \epsilon+d_2 \epsilon \delta + d_3 \epsilon \delta^2+ \dots \right)+\left(\bar{d}_1 \left(\frac{k}{\km} \right)^2 \epsilon +\bar{d}_2 \left(\frac{k}{\km} \right)^2 \epsilon \delta + \bar{d}_3 \left(\frac{k}{\km} \right)^2 \epsilon \delta^2+ \ldots \right)+   \ \ldots \ , \n
\eea
where $\ldots$ includes terms that are higher order in perturbations and terms which are suppressed by higher powers of $\frac{k}{\km}$. In the power spectrum, terms like $\epsilon \delta$ and $\epsilon \delta^2$ are degenerate with the contribution of the constant stochastic correlation function $\langle \epsilon^2 \rangle$:
\bea
&&\langle \delta_h^{(\epsilon)}\delta_h^{(\epsilon)}\rangle = d_1^2 \langle \epsilon^2 \rangle +d_2^2 \langle [\epsilon \delta]^2 \rangle+ d_1 d_3 \langle \epsilon  [\epsilon \delta^2] \rangle+\bar{d_1}d_1 \left(\frac{k}{\km} \right)^2 \langle \epsilon^2 \rangle \n
&& \qquad  \qquad + \bar{d_2}d_2 \left(\frac{k}{\km} \right)^2 \langle  [\epsilon \delta]^2 \rangle
+ \bar{d_3}d_1 \left(\frac{k}{\km} \right)^2 \langle \epsilon [\epsilon \delta^2] \rangle + \ \ldots \n
&&=  \langle \epsilon^2 \rangle \left( d_1^2 +(d_2^2+d_1 d_3)\int^{\Lambda_{UV}} d^3q P_{11}(q)+(d_2\bar{d}_2+d_1 \bar{d}_3) \left(\frac{k}{\km} \right)^2 \int^{\Lambda_{UV}} d^3q P_{11}(q) + \ \ldots   \right) \ . \n
\eea
The factor $\int^{\Lambda_{UV}} d^3q P_{11}(q)$ is a potentially large number that depends on the UV cutoff of the theory, $\Lambda_{UV}$, but this $\Lambda_{UV}$-dependence is absorbed by adjusting the value of $d_1$. The same is true for the higher-derivative terms, so after renormalization we have 
\bea
\langle \delta_h^{(\epsilon)} \delta_h^{(\epsilon)} \rangle_{ren} &=& d_{1, ren}^2  \langle \epsilon^2 \rangle + d_{2,ren}^2  \left(\frac{k}{\km} \right)^2  \langle \epsilon^2 \rangle + \ \ldots  \ ,
\label{densitystoch}
\eea
where we have neglected terms with higher powers of $k/\km$. Since we expect the constant stochastic term to be proportional to $\bar{n}^{-1}_W$, \eqn{densitystoch} can be written as:
\bea
\langle \delta_h^{(\epsilon)} \delta_h^{(\epsilon)}\rangle_{ren} &=&  \frac{1 }{\bar{n}_W}\left( \tilde{d}_{\epsilon,1}+\tilde{d}_{\epsilon,2}\left(\frac{k}{\km} \right)^2 + \ \ldots \right)  \ ,
\eea
where ${\tilde{d}_{\epsilon,1}}$ and $\tilde{d}_{\epsilon,2}$ are numbers that we expect to be order one. We will discuss the stochastic terms for $\theta_h$ in Section \ref{stochred} when we find the full expression for the stochastic biases in redshift space.

%
%
%
%

\section{Biased tracers in redshift space}

\subsection{Review of the EFT of halos in redshift space}\label{biasred}

The expansion of biased tracers in redshift space was derived in \cite{Senatore:2014vja}. We will review those results in this section. In the distant-observer approximation, the change of coordinates from real space to redshift space is given by
\be
\xv_r=\xv+\frac{\hat{z} \cdot \vec{v}}{a H}\hat{z} \ ,
\ee
where the line of sight is taken to be along the $z$-axis. Under a change of coordinates $\xv \to \xv_r$ the halo density field transforms as
\be
1+\delta_{h,r}(\xv_r)=\left(1+\delta_h(\xv) \right) \left|\frac{\partial \xv_r}{\partial \xv} \right|^{-1} \ ,
\ee
so in Fourier space the relation between the redshift-space halo density field $\delta_{h,r}$ and the real space halo density $\delta_h$ is
\be
\delta_{h,r}(\kv)=\delta(\xv)+\int d^3x \ e^{-i \kv \cdot \xv} \left(\exp\left(- i \frac{k_z }{a H} v_{h,z}(\xv)\right)-1 \right) \left(1+\delta_h(\xv) \right) \ .
\label{rsdef}
\ee

In the Eulerian approach this expression is Taylor expanded order by order in the fields $\delta_h$ and $v_h^i$. This expansion does not correctly treat the effects of long wavelength displacements, but this will be corrected by the IR resummation procedure described in Section \ref{secir}. The Taylor expansion of \eqn{rsdef} up to cubic order is
\bea
\delta_{h,r}(\kv)&=&\delta(\kv)-i \frac{k_z}{aH}v_{h,z}(\kv)+\frac{i^2}{2}\left(\frac{k_z}{a H} \right)^2[v_{h,z}^2]_{\kv}-\frac{i^3}{3!}\left(\frac{k_z}{a H} \right)^3[v_{h,z}^3]_{\kv}-i \frac{k_z}{a H}[v_{h,z} \delta_h]_{\vec{k}} \n
&&+\frac{i^2}{2}\left(\frac{k_z}{a H} \right)^2[v_{h,z}^2\delta_h]_{\kv} \ ,
\label{taylor}
\eea
where $[\ldots]_{\kv}$ represents the Fourier transform of the quantity in brackets \cite{Senatore:2014vja}. The terms $[v_{h,z}^2]_{\kv}$, $[v_{h,z}^3]_{\kv}$, $[v_{h,z} \delta_h]_{\vec{k}}$, and $[v_{h,z}^2\delta_h]_{\kv}$ must be renormalized because the product of two fields at the same location depends on UV modes in an uncontrolled manner. Since redshift space is simply a change of coordinates from real space, so far the expansion for $\delta_h$ in redshift space is the same as it was for the dark matter field \cite{Senatore:2014vja}. The only subtlety is in these contact terms, which arise because the change of coordinates involves products of fields at coincidence. In the case of the dark matter density, the renormalization for the contact operator $[v_z \delta]$ cancels with the renormalization of the linear velocity field because together they form the momentum $\pi_z$. Due to the continuity equation, $\pi_z$ is already renormalized by the counter-terms for $\delta$ \cite{Senatore:2014vja}. In the case of halos, we no longer have conservation of mass or momentum, so this argument does not apply and we need to renormalize each operator separately. This means that we have one additional contact term with respect to those of dark matter that must be renormalized, $[v_{h,z} \delta_h]$.

To renormalize the contact terms, we will write all terms in $\delta_h$ and $v^i_h$ that have the same transformation properties as the contact terms under Galilean transformations, to lowest order in derivatives. After simplifying using the linear equations of motion, the renormalized contact terms are \cite{Senatore:2014vja}:
\bea
[v_{h,z}\delta_h]_{\vec{k},r} &=& [v_{h,z}\delta_h]_{\kv}+i c_{r,4}\frac{a H}{\km}\frac{\kz}{\km} \delh^{(1)}+ {\rm stoch.} \n
\left[v_{h,z}^2 \right]_{\vec{k},r} &=& \left [v_{h,z}^2 \right]_{\kv}+ \left(\frac{a H}{\km} \right)^2c_{r,2} \delta^{(1)}+ \left(\frac{a H}{\km} \right)^2 \left(\frac{\kz}{k} \right)^2 c_{r,3} \delta^{(1)} +{\rm stoch.}  \n
\left[v_{h,z}^3 \right]_{\kv,r}  &=&  \left[v_{h,z}^3 \right]_{\kv}+3 \left(\frac{a H}{\km} \right)^2 c_{r,1} v^{(1)}_{z} +  {\rm stoch.}  \n
\left[v_{h,z}^2 \delh \right]_{\kv,r} &=&  \left[v_{h,z}^2 \delh \right]_{\kv} +\left(\frac{a H}{\km} \right)^2 {c_{r,5} } \delh^{(1)}  + {\rm stoch.} 
\label{contact}
\eea
Notice that the counter-terms of $\left[v_{h,z}^2 \right]_{\kv,r}$ and $\left[v_{h,z}^3 \right]_{\kv,r}$ are proportional to $\delta^{(1)}$, not $\delta_h^{(1)}$,  because due to the equivalence principle, they must be equal to $\left[v_{z}^2 \right]_{\kv,r}$  and $\left[v_{z}^3 \right]_{\kv,r} $ respectively, to leading order in derivatives. This means that the parameters $c_{r,1}$ and $c_{r,2}$ are equal to the corresponding parameters for dark matter. In addition, notice that the response of $\left[v_{h,z}^2 \delh \right]_{\kv,r}$ is proportional to a different parameter than the response of $\left[v_{h,z}^3 \right]_{\kv,r}$, which was not realized in \cite{Senatore:2014vja}. Indeed, $c_{r,5}$ parameterizes also the response to $\delta_h$, which will depend on halo population, while $c_{r,1}$ only depends on the dark matter velocity.

Since the vorticity is negligible at this order in perturbation theory, we can rewrite the velocity field in terms of $\theta_h$. Using the definition $v_{h,z}=-a H f  \frac{\partial_z}{\partial^2}\theta_h$, \eqn{taylor} becomes
\bea
&&\delhr =\delh+ f \left(\frac{\kz}{k} \right)^2 \theta_h \n
&&+i \kz f \left [\dzdsq \theta_h \delh \right ]_{\kv}-\frac{1}{2}\kz^2 f^2 \left [\dzdsq \theta_h \dzdsq \theta_h \right]_{\kv} -\frac{i}{6} \kz^3 f^3 \left[ \dzdsq \theta_h  \dzdsq \theta_h \dzdsq \theta_h \right]_{\kv} -\frac{1}{2}\kz^2 f^2 \left[ \dzdsq \theta_h  \dzdsq \theta_h \delh \right]_{\kv} \n
&&+\left(\frac{\kz}{\km}\right)^2 \left( c_{r,4} \delh^{(1)} -\frac{1}{2}c_{r,2} \delta^{(1)}-\frac{1}{2}  \left(\frac{\kz}{k} \right)^2 c_{r,3} \delta^{(1)}+\frac{1}{2} c_{r,1} f \left(\frac{\kz}{k} \right)^2 \delta^{(1)} -\frac{1}{2} { c_{r,5} } \delh^{(1)} \right) \n
&& + \  \delta_{\rm stoch}  + \ \ldots \ \ ,
\label{brackets}
\eea
where the third line contains the counter-terms generated in the renormalization of the contact terms in the second line and $\delta_{\rm stoch}$ refers to the stochastic terms generated by the renormalization, which we will discuss in the next section.

From the first line of \eqn{brackets}, we see that when we use \eqn{exppert} to substitute in for $\delta_h$ and $\theta_h$, we find the additional counter-term 
\be
c_{\rm ct}^{(\delta_h)}\delta^{(3,{\rm ct})}+f \left(\frac{\kz}{k} \right)^2 c_{\rm ct}^{(\theta_h)}\delta^{(3,{\rm ct})} \ ,
\ee
where $\delta^{(3,{\rm ct})}=(k^2/k_{\rm NL}^2)\delta^{(1)}$ is the counter-term for dark-matter and we have used the notation $A=\{\delta_h,\theta_h\}$. Thus the full counter-term in redshift space is given in terms of the linear dark matter density  as:
\bea
\delhr^{ (3,{\rm ct})} &=&\left(c^{(\delta_h)}_{\rm ct}+f \mu^2 c^{(\theta_h)}_{\rm ct} \right) \frac{k^2}{k_{\rm NL}^2} \delta^{(1)}+\frac{1}{2}\left(c_{r,1}f- c_{r,3}\right) \mu^4 \left(\frac{k}{\km}\right)^2  \delta^{(1)} \n
&&+\left(\left( c_{r,4}-\frac{1}{2} { c_{r,5}} \right) K_{\delta_h}^{(1)}-\frac{1}{2}c_{r,2}\right)\mu^2 \left(\frac{k}{\km}\right)^2 \delta^{(1)} \ ,
\eea
where we have defined $\mu= k_z/k$.

This expression simplifies to only three independent counter-terms, one from the biased dark matter counter-term and two from the transformation to redshift space:
\be
\delhr^{(3, {\rm ct})}=c^{(\delta)}_{\rm ct} \frac{k^2}{k_{\rm NL}^2} \delta^{(1)}+\tilde c_{r,1}\mu^2 \left(\frac{k}{\km}\right)^2  \delta^{(1)}+ \tilde c_{r,2}\mu^4 \left(\frac{k}{\km}\right)^2  \delta^{(1)} \ ,
\ee
where the new counter-term parameters $\tilde{c}_{r,1}$ and $\tilde{c}_{r,1}$ are given in terms of the original ones as
\bea
\tilde{c}_{r,1} &\equiv&\left( c_{r,4}-\frac{1}{2}{ c_{r,5}} \right) b_1-\frac{1}{2}c_{r,2}+f c_{\rm ct}^{(\theta_h)} \left(\frac{\km}{k_{\rm NL}} \right)^2 \n
\tilde{c}_{r,2} &\equiv& \frac{1}{2} \left( f c_{r,1}-c_{r,3} \right)\ .
\eea
Notice that since $\tilde{c}_{r,2}$ does not contain a bias coefficient, it is equal to the corresponding parameter for dark matter. Thus we only need one additional parameter with respect to the dark matter to describe biased tracers in redshift space, excluding stochastic terms which we will describe in the next section.

%
%
%
%
%

\subsection{Stochastic halo bias in redshift space}\label{stochred}

Now we turn to the stochastic terms for the halo power spectrum in redshift space. One contribution to the stochastic terms comes when we substitute the real-space halo stochastic terms in the first line of \eqn{brackets}, i.e.
\be
 \delta_{h,r}^{(\epsilon)}= \delta_h^{(\epsilon)}+f \mu^2 \theta_{h}^{(\epsilon)} \ .
\ee

We previously discussed the stochastic terms for $\delta_h$ in Section \ref{stochhalo}, but we still need to find the stochastic terms for $\theta_h$. Recall that diffeomorphism invariance requires all the bias terms for $v_h^i$ to be derivative-suppressed. This argument also applies to the stochastic terms because in the rest frame of the dark matter, the halo simply inherits the velocity of the dark matter in each realization. Therefore the $k \to 0$ limit of the stochastic terms for the velocity of halos is the same as that for the dark matter, and thus $v_h^i$ cannot include any constant stochastic terms because the stochastic terms of the dark matter velocity are already derivative-suppressed. This means that the leading stochastic term in $v_h^i$ goes like $\partial_i \epsilon$.

Since we are working with the velocity divergence, we get one additional derivative, and so the stochastic expansion for $\theta_h$ starts at order $k^2$:
\be
\theta_h^{(\epsilon)}=\bar{c}^2_{1,ren} \left( \frac{k}{\km} \right)^2 \epsilon + \  \ldots \ .
\ee
From \eqn{densitystoch}, we can express the stochastic halo density in terms of renormalized coefficients as
\be
\delta_h^{(\epsilon)}=d_{1,ren} \epsilon +  d_{2, ren} \left( \frac{k}{\km} \right)^2\epsilon+ \  \ldots \  ,
\ee
so the resulting stochastic terms in redshift space are 
\be
\delta_{h,r}^{(\epsilon)}={d}^2_{1,ren} \epsilon +  ({d}^2_{2,ren}+f \mu^2 \bar{d}^2_{1,ren}) \left( \frac{k}{\km} \right)^2\epsilon+ \  \ldots \   .
\label{firststoch}
\ee

We also need to consider the stochastic terms due to the renormalization of the contact terms in the transformation to redshift space, which are represented as $\delta_{\rm stoch}$ in \eqn{brackets}. From \eqn{taylor}, we see that $\left[v_{h,z}^3\right]_{\kv,r}$ comes into $\delta_{\rm stoch}$ with three derivatives, so its stochastic contribution is negligible compared to \eqn{firststoch}. The terms $\left[v_{h,z}^2  \right]_{\vec{k},r}$ and $\left[v_{h,z}^2 \delh \right]_{\kv,r}$ are multiplied by the factor $k_z^2$, so we only need to keep their constant stochastic terms, and $[v_{h,z} \delta_h]_{\vec{k},r}$  comes in with only one factor of $k_z$, so we need to keep its stochastic terms up to order $k^1$. These terms are schematically:
\bea
\hat{z}_i[v_{h}^i\delta_h]_{\vec{k},r} &=& \hat{z}_i(\epsilon^i+k^i \epsilon+ \ldots \ ) \n
\hat{z}_i\hat{z}_j\left[v_{h}^i v_{h}^j \right]_{\vec{k},r} &=& \hat{z}_i\hat{z}_j(\epsilon^{ij} + \ldots \ )  \n
\hat{z}_i \hat{z}_j\left[v_{h}^i v_{h}^j \delh \right]_{\kv,r} &=&  \hat{z}_i\hat{z}_j(\epsilon^{ij} + \ldots \ )  \ ,
\label{contactep}
\eea
where $\epsilon$, $\epsilon^i$, and $\epsilon^{ij}$ are some vector fields.
Thus the contribution to $\delta_{\rm stoch}$ to second order in derivatives goes like
\be
\delta_{\rm stoch} \sim k_z \hat{z}_i( \epsilon^i+k^i \epsilon)+ k_z^2\hat{z}_i\hat{z}_j \epsilon^{ij} \ .
\ee

In the power spectrum,  $\delta_{\rm stoch}$ can correlate with both itself and with the other stochastic terms in \eqn{firststoch}. When  $\delta_{\rm stoch}$ contracts with $\delta_{h,r}^{(\epsilon)}$, we find the following terms up to order $k^2$:
\be
\langle \delta^{(\epsilon)}_{h,r}  \delta_{\rm stoch} \rangle \sim  \mu k \hat{z}_i \left( \langle  \epsilon^i \epsilon \rangle+k^i \langle \epsilon^2 \rangle \right)+ \mu^2 k^2\hat{z}_i\hat{z}_j \langle \epsilon^{ij} \epsilon \rangle\ .
\label{crosstoch}
\ee
Before they are projected on the $z$-axis, the correlation functions $ \langle \epsilon^i \epsilon \rangle$ and $ \langle \epsilon^{ij} \epsilon \rangle$ must be Lorentz-invariant. Thus, $ \langle \epsilon^{ij} \epsilon \rangle$ must be proportional to $\delta^{ij}$, and since the only vector with one index that we can write down is $k^i$,  $ \langle \epsilon^i \epsilon \rangle$ must be proportional to  $k^i \langle \epsilon^2 \rangle $. This means that \eqn{crosstoch} takes the form:
\be
\langle \delta^{(\epsilon)}_{h,r}  \delta_{\rm stoch} \rangle \sim  \mu k \hat{z}_i k^i \langle \epsilon^2 \rangle + \mu^2 k^2\hat{z}_i\hat{z}_j \delta^{ij} \langle \epsilon^2 \rangle  \sim \mu^2 k^2 \langle \epsilon^2 \rangle \ .
\label{cross}
\ee
Similarly, when contracted with itself, $\delta_{\rm stoch}$ gives the term:
\be
\langle \delta_{\rm stoch}^2 \rangle \sim k_z^2 \hat{z}_i\hat{z}_j  \langle \epsilon^i \epsilon^j \rangle \sim k_z \hat{z}_i\hat{z}_j \delta^{ij} \langle \epsilon^2 \rangle \sim \mu^2 k^2 \langle \epsilon^2 \rangle \ ,
\ee
which is the same as what we found in \eqn{cross}. Both of these terms are degenerate with the contribution to the power spectrum from \eqn{firststoch}. Thus all of the stochastic terms in redshift space due to the renormalization of the contact terms are degenerate with the contributions from the halo stochastic biases up to order $k^2$ in the power spectrum.

This means we can write the stochastic halo power spectrum in redshift space up to order $k^2$ in terms of only three independent parameters,
\be
\langle \delta_{h,r}\delta_{h,r} \rangle_\epsilon =\frac{1}{\bar{n}_W} \left(c_{\epsilon,1} + c_{\epsilon,2} \left( \frac{k}{\km} \right)^2 +c_{\epsilon,3} f\mu^2 \left( \frac{k}{\km} \right)^2  \right) \ ,
\label{stochfin}
\ee
and these are the parameters that we will use to fit to simulations. Notice that since the $c_{\epsilon,i}$ are dimensionless and expected to be order one, the overall size of the stochastic counter-term is set by the mean squared halo density in \eqn{meansq}, which will determine how many stochastic terms in the derivative expansion need to be included along with the other counter-terms in the fits. We will see in Section \ref{secfits} that all three terms in \eqn{stochfin} will be needed and that the $k^4$ terms are indeed negligible.

\subsection{Halo-halo power spectrum in redshift space}

Now we turn back to the expansion for the contact terms in \eqn{brackets}. When we collect the contact terms order by order, we have
\bea
\delhr^{(1)}(\kv)&=&\delh^{(1)}+ f \left(\frac{\kz}{k} \right)^2 \theta_h^{(1)} \n
\delhr^{(2)}(\kv)&=&\delh^{(2)}+ f \left(\frac{\kz}{k} \right)^2 \theta_h^{(2)} +i\kz f \delta_{[\frac{\partial_z}{\partial^2} \theta_h \delh]}^{(2)} (\kv)-\frac{1}{2}\kz^2 f^2\delta_{[\frac{\partial_z}{\partial^2} \theta_h \frac{\partial_z}{\partial^2} \theta_h]}^{(2)} (\kv) \n
\delhr^{(3)}(\kv)&=&\delh^{(3)}+ f \left(\frac{\kz}{k} \right)^2 \theta_h^{(3)} +i\kz f \delta_{[\frac{\partial_z}{\partial^2} \theta_h \delh]}^{(3)} (\kv) -\frac{1}{2}\kz^2 f^2\delta_{[\frac{\partial_z}{\partial^2} \theta_h \frac{\partial_z}{\partial^2} \theta_h]}^{(3)} (\kv) \n
&& \qquad -\frac{i}{6} \kz^3 f^3\delta_{[\frac{\partial_z}{\partial^2} \theta_h \frac{\partial_z}{\partial^2} \theta_h \frac{\partial_z}{\partial^2} \theta_h]}^{(3)} (\kv)-\frac{1}{2}\kz^2f^2 \delta_{[\frac{\partial_z}{\partial^2} \theta_h \frac{\partial_z}{\partial^2} \theta_h  \delh]}^{(3)} (\kv)  \ ,
\label{collect}
\eea
where the expressions for the $\delta^{(n)}_{[\ldots]}$ are given in \eqn{rsec} and \eqn{rthird} of Appendix \ref{rskernels}. After substituting the expressions for $\theta_h$ and $\delta_h$ from \eqn{hkernel}, the redshift-space fields will also be given in terms of integrals of $\delta^{(1)}$ with new momentum kernels defined by
\be
\delta^{(n)}_{h,r}(\kv)=\int d^3 q_1 \ldots d^3 q_n  K^{(n)}_{h,r}(\q_1, \ldots , \q_n)_{\sym} \delta^3_D(\kv-\q_1 \ldots - \q_n) \delta^{(1)}(\q_1) \ldots \delta^{(1)}(\q_n) \ . 
\ee
As shown in \eqn{rsec} and \eqn{rthird}, the explicit expressions for the full halo density kernels in redshift space are
\bea
K^{(1)}_{h,r} (\q_1) &=&K_{\delta_h}^{(1)} (\q_1)+f \mu^2 K_{\theta_h}^{(1)} (\q_1) = b_1+f \mu^2 \n
K^{(2)}_{h,r}(\q_1, \q_2)&=&K_{\delta_h}^{(2)}(\q_1,\q_2)+f \mu^2 K_{\theta_h}^{(2)}(\q_1,\q_2) \n
&& \qquad +\frac{1}{2} \mu f  \left( \frac{ k q_{2z}}{q_2^2}+\frac{ k q_{1z}}{q_1^2} \right) K_{\theta_h} ^{(1)} (\q_1) K_{\delta_h} ^{(1)} (\q_2)+ \frac{1}{2}\mu^2f^2\frac{k^2 q_{1z}q_{2z}}{q_1^2 q_2^2}K_{\theta_h} ^{(1)} (\q_1) K_{\theta_h} ^{(1)} (\q_2) \n
K^{(3)}_{h,r}(\q_1, \q_2, \q_3)&=&K_{\delta_h}^{(3)}(\q_1,\q_2,\q_3)+f \mu^2 K_{\theta_h}^{(3)}(\q_1,\q_2,\q_3)  \n
&&+ \mu f \left(  \frac{k q_{3z}}{q_3^2}  \right) K^{(2)}_{\delta_h} (\qv_1,\qv_2)K^{(1)}_{\theta_h} (\q_3)  +\mu f  \left(  \frac{k(q_{1z}+ q_{2z})}{(\qv_1+\qv_2)^2}  \right)  K^{(2)}_{\theta_h} (\qv_1,\qv_2)K^{(1)}_{\delta_h}  (\q_3)\ \n
&& +\frac{1}{2}\mu^2 f^2 \left( \frac{k q_{1z}}{q_1^2}\frac{k q_{2z}}{q_2^2} \right) K^{(1)}_{\theta_h} (\q_1) K^{(1)}_{\theta_h} (\q_2) K^{(1)}_{\delta_h}  (\q_3) \n
&&  + \mu^2 f^2  \left(  \frac{k(q_{1z}+ q_{2z})}{(\qv_1+\qv_2)^2} \frac{k q_{3z}}{q_3^2} \right) K^{(2)}_{\theta_h} (\qv_1,\qv_2)K^{(1)}_{\theta_h} (\q_3) \n
&&+\frac{1}{6}\mu^3 f^3 \left( \frac{k q_{1z}}{q_1^2}\frac{k q_{2z}}{q_2^2}\frac{k q_{3z}}{q_3^2} \right) K^{(1)}_{\theta_h}  (\q_1) K^{(1)}_{\theta_h}  (\q_2) K^{(1)}_{\theta_h}  (\q_3) \ , 
\label{allredshift}
\eea
where the $K_{A}^{(n)}$ are the kernels for halo species $A$ given in \eqn{hkernels} of Appendix \ref{sec3}, and we have used the notation $A=\{\delta_h,\theta_h\}$. Using these kernels, we can now compute the halo power spectrum in redshift space,
\bea
\langle \delhr(\kv) \delhr(\kv) \rangle &=& \langle \delhr^{(1)}\delhr^{(1)} \rangle+ \langle \delhr^{(2)} \delhr^{(2)} \rangle+ 2 \langle \delhr^{(1)} \delhr^{(3)} \rangle +\langle \delhr \delhr \rangle_{\rm ct}+ \langle \delhr \delhr\rangle_{\epsilon}\n
&=& \bigl(K^{(1)}_{h,r} \bigl)^2 P_{11}(k)+2 \int d^3 \q \   \left(K^{(2)}_{h,r}(\q,\kv-\q)_{\sym} \right)^2 P_{11}(|\kv-\q|)P_{11}(q) \n
&&+6\int d^3 \q  \ K^{(3)}_{h,r}(\q,-\q, \kv)_{\sym} K^{(1)}_{h,r} P_{11}(q) P_{11}(k) + \langle \delhr \delhr \rangle_{\rm ct}+ \langle \delhr \delhr \rangle_{\epsilon} \ . \n
\label{finalpsred}
\eea
The contribution from the counter-terms is:
\bea
\langle \delhr(\kv) \delhr(\kv) \rangle_{\rm ct} &=& 2 \langle \delhr^{(1)}(\kv){\delhr^{(3,{\rm ct})}}(\kv) \rangle \n
&=&2P_{11}(k)\left(K_{\delta_h} ^{(1)}+f\mu^2 K_{\theta_h} ^{(1)} \right) 
\left(  \mu^2 \left(\frac{k}{\km} \right)^2 \tilde{c}_{r,1}+ \mu^4 \left(\frac{k}{\km} \right)^2 \tilde{c}_{r,2}  +c_{\rm ct}^{(\delta_h)}  \left(\frac{k}{k_{\rm NL}} \right)^2 \right) \n
&=&2 P_{11}(k) (b_1+f \mu^2)\left(  \mu^2 \left(\frac{k}{\km} \right)^2 \tilde{c}_{r,1}+ \mu^4 \left(\frac{k}{\km} \right)^2 \tilde{c}_{r,2}  +c_{\rm ct}^{(\delta_h)} \left(\frac{k}{k_{\rm NL}} \right)^2 \right) 
\ ,
\label{finalct}
\eea
and the contribution from the stochastic terms is given in \eqn{stochfin}.

%
%
%
%

\section{IR resummation}\label{secir}
So far this calculation has been done in a fixed Eulerian frame defined by the coordinates $\xv$, rather than in the Lagrangian frame following the fluid particles themselves. This means that we have expanded perturbatively in all of the tidal forces and displacements, which are controlled by the following parameters:
\bea
\epsilon_{s>} &=& k^2 \int_k^{\infty} d^3 q \ \frac{P_{11}(q)}{q^2} \n
\epsilon_{s<} &=& k^2 \int^k d^3 q \ \frac{P_{11}(q)}{q^2} \n
\epsilon_{\delta<} &=& \int^k d^3q  \ P_{11}(q) \ .
\label{params}
\eea
$\epsilon_{s>}$ parameterizes the effect of displacements due to momenta larger than $k$, $\epsilon_{\delta<}$ controls the tidal forces due to momenta smaller than $k$, and $\epsilon_{s<}$ parameterizes the effect of long-wavelength displacements. Notice that $\epsilon_{\delta>}$, which parameterizes the effect of the tidal forces due to momenta greater than $k$, does not appear.

The Eulerian expansion assumes that all of the parameters \eqn{params} are small. This is valid for both $\epsilon_{s>}$ and $\epsilon_{\delta<}$ because they both arise in the loops and are proportional to powers of $k/k_{M}$. However, the final parameter $\epsilon_{s<}$ is not generically small. Although it is expected to cancel in equal-time correlators due to the equivalence principle because both fields have undergone the same constant drift \cite{Scoccimarro:1995if}, this is no longer true in the presence of the BAO oscillations because displacements between the BAO scale and the nonlinear scale do not cancel \cite{Senatore:2014via}. Also, in non-equal-time correlators, and in correlators where there is a relative velocity between species that cannot be transformed away, such as the one between baryons and dark matter \cite{Lewandowski:2014rca}, $\epsilon_{s<}$ is generically order one to begin with.

The Eulerian approach does not correctly take into account the effect of long-wavelength displacements because they can accumulate over time along the fluid flow if there is a large bulk velocity. The remedy is to resum the non-perturbative effects of the linear displacement power spectrum, as described in \cite{Senatore:2014via}. This IR-resummation method makes use of the Lagrangian approach, which tracks the displacement of particles from their initial position rather than their absolute position in time. In this approach, described in the context of the EFT of LSS in  \cite{Porto:2013qua}, $\epsilon_{s<}$ is automatically small because relative displacements are measured in coordinates that are co-moving with the fluid, so any large displacements caused by the motion of the fluid as a whole do not contribute to correlation functions. The IR resummation procedure corrects the Eulerian power spectra by convolving them with terms that account for the effects of these linear displacements of the fluid, i.e. the part of $\epsilon_{s<}$ that is due to the bulk motion.

Let us review how the resummation works in real space and then we will discuss how it changes when going to redshift space. In Lagrangian space, the correlation function for the density is related to the correlation functions of the displacements $s^i$ from the initial coordinates $q^i$, 
\be
P(k)= \int d^3q \  e^{-i \kv \cdot \qv} \langle e^{-i \kv \cdot(\vec{s}(\qv,t)-\vec{s}(0,t))} \rangle \ .
\label{eqlag}
\ee
If we were to assume all displacements were small and expand the exponential in \eqn{eqlag} in a Taylor series, we would recover the Eulerian power spectrum. However, we would like to keep linear displacements in the exponential because they can become large and potentially break the perturbative expansion, so instead we will expand this correlation function  in cumulants:
\be
P(k) = \int d^3q \  e^{-i \kv \cdot \qv} \  e^{\sum_{n=0}^\infty \frac{1}{n!} \langle (\kv \cdot (\vec{s}(\qv,t)-\vec{s}(0,t)))^2 \rangle} \ .
\ee

We are not able to calculate this infinite sum explicitly so we must expand to some finite order in perturbation theory, $P\vert\vert_j$, where the double bar denotes expanding up to order $j$. This entails incorrectly expanding in the large displacements. However, we can recover the correct exponential behavior of the linear displacements using the leading term in the cumulant expansion,
\be
K_0(\kv, \qv, t)=e^{-\frac{1}{2} \langle (\kv \cdot (\vec{s}(\qv,t)-\vec{s}(0,t)))^2\rangle } \ .
\label{k0def}
\ee
If we convolve the truncated spectra with the following expression in terms of $K_0$,
\be
P(k)\vert_N= \int d^3 k' \sum_{j=0}^N K_0(k) \cdot( K_0(k)^{-1})\vert\vert_{N-j} P(k')_j \ ,
\label{convolve}
\ee
we will retain the non-perturbative behavior of the linear displacements \cite{Senatore:2014via}. This is denoted by the single bar on the lefthand side of \eqn{convolve}, which represents expanding up to order $N$ in $\epsilon_{\delta<}$ and $\epsilon_{s>}$, but treating the IR displacements exactly. This procedure works because $K_0^{-1}||_{N-j}$ cancels the improper perturbative expansion that has been done in expanding $P(k)$ up to order $j$, and $K_0$ restores the exponential behavior of the linear displacements.

It was shown in \cite{Senatore:2014vja} that the IR resummation for halos is the same as the procedure for dark matter with the replacement $\delta \to \delta_h$ and $v^i \to v_h^i$. This is because the displacements are proportional to the halo velocity, which we have seen is equal to the dark matter velocity at leading order in derivatives. Thus the only change to the IR resummation in our case comes from the change of coordinates to redshift space, which is described in \cite{Senatore:2014vja} and which we will now discuss.

The key difference in redshift space is that we must treat separately the displacements parallel to and perpendicular to the line of sight due to the reduced symmetry. As a result, $K_0$ becomes a function of $\kv+f\mu k^2 \hat{z}$ instead of $\kv$. We define a new $\tilde K_0(\kv)$ for redshift space:
\be
\tilde K_0(\kv)=\exp \left[ -\frac{1}{2} \left \langle \left(\left(\kv+\mu^2 f k^2\hat{z}\right) \cdot \left(\vec{s}_h^{(1)}(\qv)-\vec{s}^{(1)}_h(\vec{0}) \right) \right)^2 \right \rangle \right] \ ,
\ee
and the calculation proceeds in the same way that it would in real space after substituting $\tilde K_0$ for $K_0$. In redshift space, it is convenient to expand the power spectra in multipole moments, so we will need to compute
\be
P_l^r(k)|_N=\sum_{j=0}^N \sum_{l'}\int \frac{dk' k'^2}{2\pi^2}M_{l,l'}\vert\vert_{N-j}(k,k')P_{l'}^r(k')_j \ ,
\ee
where $M_{l,l'}\vert\vert_{N-j}(k,k')$ is the factor $\tilde K_0\cdot ( \tilde K_0^{-1})\vert\vert_{N-j}$ written in the monopole expansion:
\bea
M_{l,l'}(k,k') \vert\vert_{N-j}&=&\int dq \  j_{l'}(k'q) i^{l'} q^2 \frac{2l+1}{2}\int_{-1}^1 d \mu \int d^2 \hat{q} \ e^{-i \qv \cdot \kv} \tilde K_0(\kv)\cdot ( \tilde K_0(\kv)^{-1} )\vert\vert_{N-j} \n
&& \qquad \qquad \qquad \qquad \times\mathcal{P}_l(\mu) \mathcal{P}_{l'}(q_z/q) \ ,
\label{meq}
\eea
 and where the $j_l(x)$ are the first-order Bessel functions and the $\mathcal{P}_l(x)$ are the Legendre polynomials.

The details of this calculation can be found in \cite{Senatore:2014vja} and \cite{Lewandowski:2015ziq}. The real complication for the IR resummation in redshift space is that since $\tilde K_0$ now depends on the angular coordinate $\mu$,  there is an additional integral that must be done. This makes the numerical integrals much more difficult. However, a modified procedure was developed in \cite{Lewandowski:2015ziq}, in which a controlled expansion of the exponent in $\tilde K_0$ is performed to reduce the computational load. We will implement this procedure. The explicit expressions we used for the resummation of the halo power spectra are given in Appendix \ref{irResum}.

%
%
%
%

\section{Fits to simulations}\label{secfits}

Using \eqn{finalpsred} and the IR resummation procedure in \eqn{irresum} of Appendix \ref{irResum}, we can now calculate the EFT power spectrum for generic biased tracers in redshift space and compare the results to simulations. Here we compare the redshift-space power spectra to halo power spectra measured from one of the {\em Dark Sky} simulations \cite{Skillman:2014qca}.  The {\em Dark Sky} simulation used herein is a 1 $h^{-1} {\rm Gpc}$ box simulated with $10240^3$ particles, with cosmological parameters $\Omega_m=0.295$, $\Omega_\Lambda=0.705$, $H_0=68.8 {\rm \ km \cdot s^{-1} Mpc^{-1}}$, and $\sigma_8=0.83$.  This was run with the 2HOT code of Warren et al~\cite{Warren:2013vma}. The {\tt Rockstar} halo finder \cite{Behroozi:2011ju} was used to identify halos. This halo finder was run on a downsample of the full simulation, that contains 1/32 of the total particle number (see~\cite{Jennings:2015lea, Lehmann:2015ioa} for further details). The power spectra of these halos was measured as described in Jennings et al.~\cite{Jennings:2015lea}.  Here we specifically use the power spectra of all halos with masses of $M_{200} > 1\times 10^{11}$ $h^{-1}\Msun$ at $z\sim0.67$, with a number density $\bar n=3.8 \cdot 10^{-2}(\hinvMpc)^3$.

Later in Sec.~\ref{app:galaxies}, we perform the same fit to a different sample. This is the $v_{m_{peak}}$ model of LRGs from \cite{Jennings:2015lea}, which has a number density $\bar n=3.9 \cdot 10^{-4}(\hinvMpc)^3$. Though this sample has a lower number density and an higher bias, a fact that could lead to a decrease in the $k$-reach of the theory at a given number with a given number of counterterms~\cite{Fujita:2016dne}, we find that the performance of the theory is comparable in the two samples (even though the cosmic variance error bars for the quadrupole are in this case a factor of two larger). This result is not surprising from the EFTofLSS point of view, as different populations, even real galaxies, represent just different UV models which, in the formalism of the EFTofLSS, are just different biased tracers described by the same set of equations, just with different coefficients.

The final IR-resummed halo power spectrum in redshift space has four bias parameters $\{ b_1$, $b_2$, $b_3$, $b_4 \}$, three ``speed of sound" parameters $\{c_{\rm ct}^{(\delta)}$, $\tilde{c}_{r,1}$, $\tilde{c}_{r,2} \} $, and three stochastic parameters $\{ c_{\epsilon,1}$, $c_{\epsilon,2}$, $c_{\epsilon,2} \}$, for a total of ten free parameters. All of these terms are dimensionless and expected to be order one. From \eqn{stochfin}, we know that the stochastic terms are multiplied by the dimensionful quantity $\bar{n}^{-1}_W $, which, for the sample $M_{200} > 1\times 10^{11}$ $h^{-1}\Msun$ of about $4\cdot 10^7$ halos, is $\bar{n}^{-1}_W \sim 105 \ (\unitskinv)^3$. Here the subscript ${}_W$ refers to the fact that the number density is estimated taking into account the width of the bin in mass and how the different masses contribute to the power spectrum.

We can now proceed to the fits. We expand the power spectrum in multipoles and fit to the power spectra for the real-space ($\mu$=0) mode, the $l=0$ mode, and the $l=2$ mode from the simulations. We add a systematic error of one percent of the $P_{l=0}$ mode to each power spectrum. There is a larger overall error for $P_{l=2}$ because it is normalized by $2l+1$. The procedure for determining the reach of the EFT fit is as follows, based on the approach of \cite{Foreman:2015lca}. A non-linear fit of the EFT power spectra with ten free parameters to the power spectra obtained in simulations is performed simultaneously for $P_{\rm real}$, $P_0$, and $P_2$ up to a given $k_{\maxx}$. This is repeated for different values of $k_{\maxx}$, and then the value of each parameter obtained for a given $k_{\maxx}$ is plotted against $k_{\maxx}$. This is shown in Fig.~\ref{bplots1h} and Fig~\ref{bplots2h} in Appendix \ref{haloplots} for the halos. 

\begin{figure}[htb!]
\centering
\includegraphics[width=9.5cm]{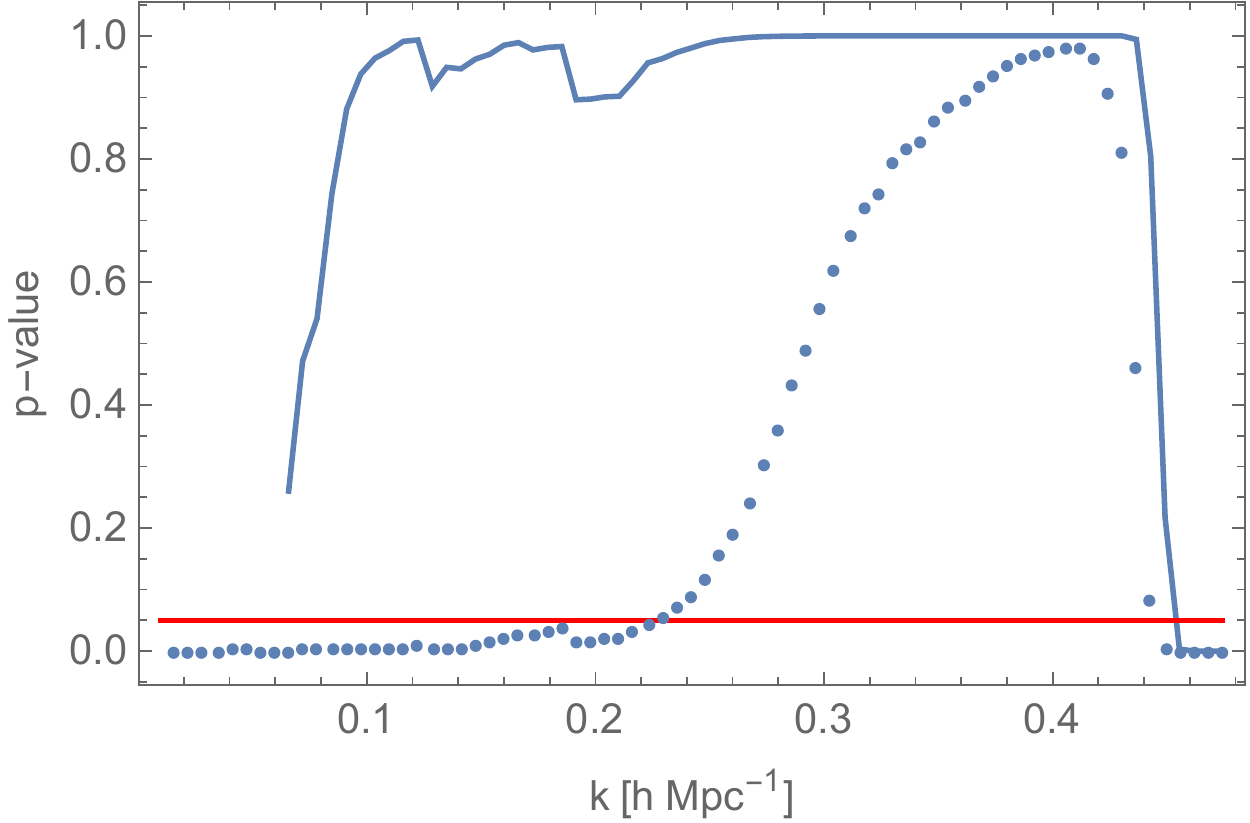}
\caption{Plot of $p$-values calculated up to a given $k$ for the IR-resummed fit depicted in Fig.~\ref{fits} with $k_{\rm fit}=0.39 \ \unitsk$.  The solid blue curve shows the $p$-value, neglecting the data points with $k<0.06 \ {\rm h \ Mpc^{-1}}$, and the dotted blue curve includes all of the low-$k$ points. The horizontal red line shows $p=0.05$.}\label{pvalueplot}
\end{figure}

The determination of the parameters will continue to improve as more points are included in a higher $k_{\maxx}$. However, at some point the value of the parameters obtained from the fit at a certain $k_{\maxx}^*$ may become incompatible with the values from the previous fits. We interpret this as being due to overfitting, and suggests that we should not fit beyond this critical $k_{\maxx}^*$, which we will label as $k_{\rm fit}$. Using this procedure, $k_{\rm fit}$ is determined as the value of $k_{\maxx}$ where any one fitting parameter becomes more than $2\sigma$ discrepant from its lower-$k$ values.  We see in Fig. \ref{bplots2h} that the values of the parameter $b_1$ begins to be inconsistent with the previous values at $k_{\maxx}=0.39 \ \unitsk$. This is the first parameter to fail, so we use this value for $k_{\rm fit}$.

 A plot of the $p$-values of the fits up to different values of $k$, shown in Fig.~\ref{pvalueplot}, confirms the goodness of fit up to $k=0.43 \  \unitsk$. In Fig.~\ref{pvalueplot}, the dashed line shows the $p$-value of the fit including all of the points measured in simulations up to $k_{\rm fit}$, and the solid line shows the $p$-value excluding the points with $k<0.06 \ {\rm h \ Mpc^{-1}}$. These  low-$k$ points cannot be well-fit by the parameters so they reduce the $p$-value until many higher-$k$ points are included. This may be due to the fact that since we are looking at a finite region of space, all our integrals in $k$ should really be sums over discrete $k$-modes, or it may be due to the large error at these low wavenumbers. We do not investigate it further as these issues affect quite long wavenumbers, so the EFT is expected to work very well.

 \begin{figure}
  \centering
  \includegraphics[width=9.5 cm]{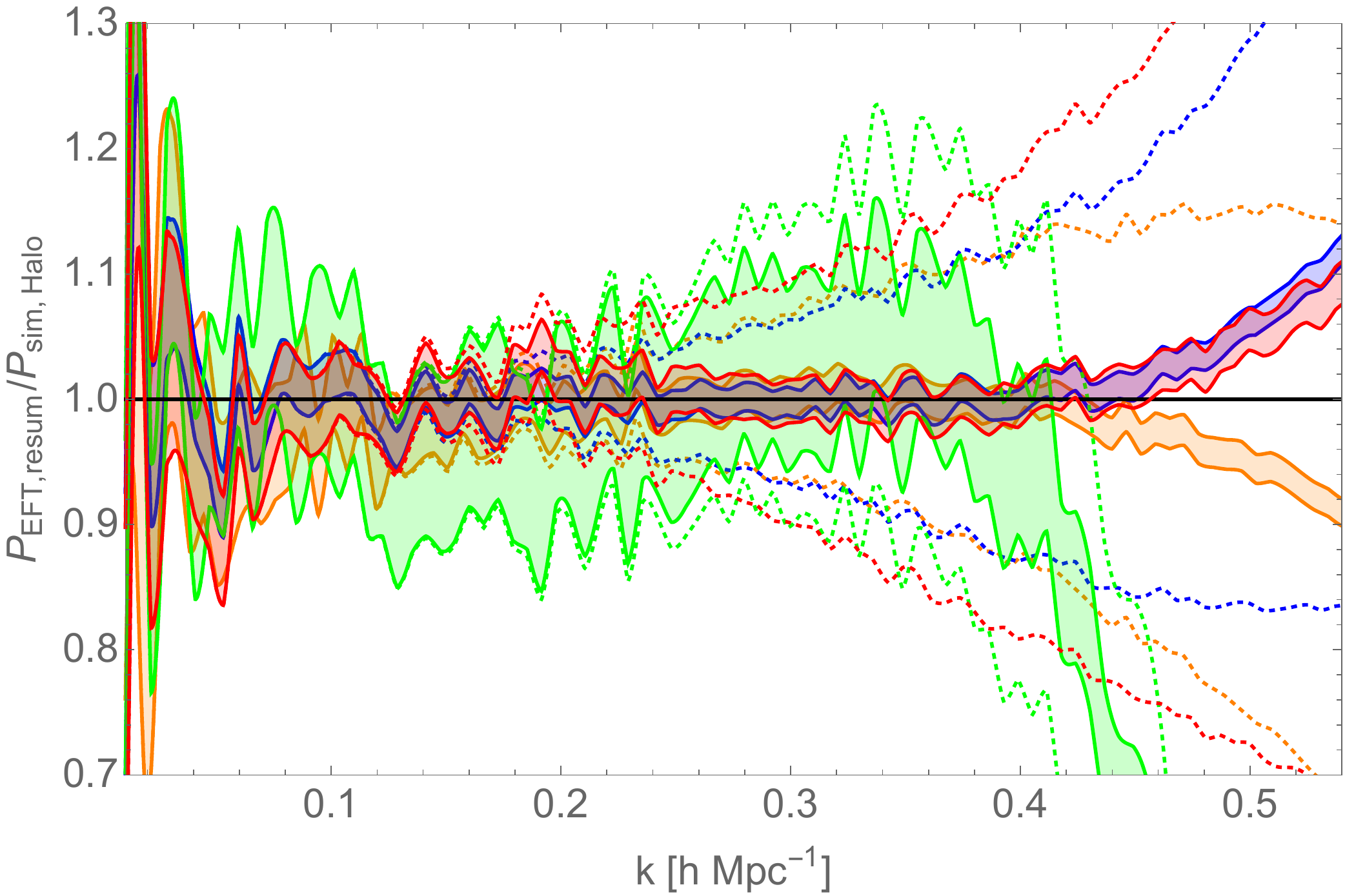}
  \caption{ \footnotesize  Results of the fits of the IR-resummed EFT power spectra at $z=0.67$ to the power spectra of halos and dark matter extracted from simulations.  The halos have masses of $M_{200} > 1\times 10^{11}$ $h^{-1}\Msun$, with a number density $\bar n=3.8 \cdot 10^{-2}(\hinvMpc)^3$. The fits were performed in the $k$-range $k_{\rm min}=0.01 \ \unitsk$ to $k_{\rm fit}=0.39 \ \unitsk$ and resulted in the best-fit parameters $\{ b_1=0.98 \pm 0.01, \ b_2=0.01\pm 2.73, \ b_3=-0.62 \pm 1.43, \ b_4=0.58 \pm 2.33, \ c_{\rm ct}^{(\delta_h)}=(5.3\pm 4.7) \left( k_{\rm NL} \ h^{-1}  {\rm Mpc} \right)^{2} , \ \tilde{c}_{r,1}=(-14 \pm 5)  \left( \km  \ h^{-1} {\rm Mpc} \right)^{2}, \ \tilde{c}_{r,2}=(-0.69 \pm 1.67 )  \left( \km  \ h^{-1}  {\rm Mpc} \right)^{2}, \ c_{\epsilon,1}=0.76 \pm 14.74, \ c_{\epsilon,2}=(8.9 \pm 3.4)  \left( \km  \ h^{-1}  {\rm Mpc} \right)^{2}, \ c_{\epsilon,3}=(8.0 \pm 7.8 )  \left( \km  \ h^{-1}  {\rm Mpc} \right)^{2} \}$ for the halos and $c_s^2=(-0.61 \pm 0.02) \left( k_{\rm NL}  \ h^{-1} \ {\rm Mpc} \right)^{2}$ for the dark matter. The shaded regions show the 1$\sigma$ error on the simulation data, which includes the error on the halo spectra from simulations described in \cite{Feldman:1993ky} and a $1\%$ error that we add in quadrature to account for unknown systematic effects. The expected theoretical error is given by the dotted lines. }
    \label{fits}
\end{figure}

The results of the fits for the IR-resummed power spectra of halos at $k_{\rm fit}=0.39 \  \unitsk$ are given in Fig.~\ref{fits}, along with the fit to the dark matter power spectrum in real space for comparison. Together with the errors from sample variance and an estimate of a systematic error in comparing theory and simulations, we include an order-of-magnitude level estimate of the theory error associated to the higher order terms we did not include in the computation~\footnote{We estimate the theoretical error as follows. First, we fit the $z=0.67$ linear matter power spectrum from CAMB as a piecewise power law \cite{Carrasco:2013mua,Pajer:2013jj}:
\bea
P_{11}^{\rm fit}(k)=(2\pi)^3 \begin{cases}  \frac{1}{ \knl^3 } \left( \frac{ k}{  \knl } \right)^{n} & \text{for} \, \,  \, k> k_{\rm tr}  \\    \frac{1}{ \bar{k}_{\rm NL}^3 } \left( \frac{ k}{  \bar{k}_{\rm NL} } \right)^{\bar{n}}  & \text{for} \, \, \, k < k_{\rm tr} \ .   \end{cases}
\eea
Then, since the two-loop term scales approximately as $P_{\rm 2-loop}/P_{11} \sim (k/k_{\rm NL})^{2(3+n)}$, we estimate the theoretical error on the dark matter power spectrum from neglecting the two-loop terms to be of order
\be
\Delta P_{\rm 1-loop}\sim P_{\rm 2-loop} \sim 2\pi^2 P_{11}^{\rm fit}(k)\left(\frac{k}{k_{\rm NL}^i} \right)^{2(3+n^i)} \ , 
\ee
where $\{k_{\rm NL}^i,n^i\}$ equals $\{\knl,n\}$ for $k>k_{\rm tr}$ and $\{\bar{k}_{\rm NL}, \bar{n} \}$ for  $k<k_{\rm tr}$, and the factor of $2\pi^2$ approximately accounts for factors coming from integration. Since our universe does not have a true power-law spectrum and since numerical factors are hard to estimate, the estimates for the theory error should be taken at the order-of-magnitude level. 
\label{theory}}. We perform a consistency check of this fitting procedure by using a different fitting procedure that includes the estimated theoretical error in Appendix~\ref{check}, and we find consistent results.  The results of the fits to the power spectra before IR-resummation are given in Fig.~\ref{noresum} of Appendix~\ref{irResum}. There we see that the IR-resummation is essential for the fit, especially for the $l=2$ mode which has oscillations of about $20\%$ that are resummed. In Fig. \ref{fits} the fits of the EFT to the halo power spectra fail at about the same wavenumber as the fit to the dark matter power spectrum, which we expect from effective field theory.  The bias parameters determined by the fit for the IR-resummed halo power spectra along with their $1\sigma$ errors determined by the fitting procedure are~\footnote{The $k^0$ stochastic term, which is parameterized by $c_{\epsilon,1}$, must be positive because, after we subtract the UV contribution for the diagrams of the 2-2 kind as we do, it represents the induced power spectrum from modes into the non-linear regime. Thus, we have implemented the constraint $c_{\epsilon,1}\ge0$ in the fits. Since Mathematica seems to us to have difficulty converging on the fits when the $c_{\epsilon,1}\ge0$ constraint is implemented, we start the parameter values of $b_1$, $b_2$, $b_3$, and $c_{\epsilon,1}$ with the center values obtained in an unconstrained fit. $b_1$ was constrained to stay within $\pm 6\%$ of the center value, $b_2$ and $b_3$ were constrained to $\pm 320\%$, and $c_{\epsilon,1}$ was bounded above by $+1100\%$ of the center value. The remaining parameters were left unconstrained.}:
\bea
b_1 &=& 0.98\pm 0.01 \n
b_2 &=& 0.01 \pm 2.73\n
b_3 &=& -0.62 \pm 1.43\n
b_4 &=& 0.58 \pm 2.33\n
c_{\rm ct}^{(\delta_h)} &=& \left( 5.3 \pm 4.7 \right) \left( \frac{k_{\rm NL}}{ h \ {\rm Mpc}^{-1}} \right)^{2} \n
\tilde{c}_{r,1}&=& \left( -14 \pm 5  \right) \left( \frac{\km}{ h \  {\rm Mpc}^{-1}} \right)^{2}\n
\tilde{c}_{r,2} &=& \left( -0.69 \pm 1.67  \right) \left( \frac{\km}{ h \  {\rm Mpc}^{-1}} \right)^{2} \n
c_{\epsilon,1} &=& \left( 0.76 \pm 14.74 \right) \n
c_{\epsilon,2} &=& \left( 8.9 \pm 3.4 \right)  \left( \frac{\km}{ h \  {\rm Mpc}^{-1}} \right)^{2} \n
c_{\epsilon,3} &=& \left( 8.0 \pm 7.8 \right)  \left( \frac{\km}{ h \  {\rm Mpc}^{-1}} \right)^{2} \ .
\eea 

Note that the errors are quite correlated. We give the correlation matrix in Appendix~\ref{haloplots}.

It is useful to provide a rough estimate of the scale $k_{\rm M}$ suppressing the higher-derivative biases of halos. We saw in \eqn{meansq}  that the stochastic power spectrum, which renormalizes the single halo contribution, can be estimated using the halo mass function. We can estimate the size of $k_{\rm M}$ by comparing the typical size of $k_{\rm M}^{-2} P_{\rm stoch}$, a higher-derivative correction to the stochastic power spectrum, to the size of $P_{\rm stoch}$:
\be
\frac{1}{k_{\rm M}^2 }\sim \frac{\int dM \frac{dn}{dM} \frac{M^2}{\rho_b^2}\frac{1}{\bar k(M)^2}}{\int dM \frac{dn}{dM}\frac{M^2}{\rho_b^2}} \ ,
\ee 
where we have taken $\bar k(M) = 2\pi (\frac{4 \pi}{3}\frac{\rho_b}{M})^{1/3}$, the inverse size of a halo of mass $M$. This gives the rough estimate $ k_{\rm M} \sim 0.9\hinvMpc$, which makes $\tilde{c}_{r,1}$ and $\tilde{c}_{r,2}$ order $1-10$, and the $c_{\epsilon,i}$ order one. Of course this estimate should be taken at the order of magnitude level.

At this point, we should compare the size of the two-derivative stochastic terms to the size of the ``speed of sound" counter-terms to know whether it was consistent to include them. The $\tilde{c}_{r,2}$ term is the smallest ``speed of sound" counter-term and the $c_{\epsilon,3}$ term is the smallest stochastic counter-term. The ratio of these terms is approximately
\be
\frac{\bar{n}^{-1}_W f \mu^2 c_{\epsilon,3} }{\mu^4c_{\tilde{r},2} P_{11}(k) } \sim \frac{400}{P_{11}(k)},
\ee
which is order one or larger for $k>0.3$. This means that the $k^2$ stochastic terms are of the same order of magnitude as the other $k^2$ counter-terms, and must be included to be consistent. Thus, we find that it was consistent to expand up to second order in derivatives in the power counting of the stochastic term. The $k^4$ terms we neglected in the derivative expansion of both the stochastic and the ``speed of sound" counter-term expressions are suppressed with respect to the ones we have kept, but may become relevant at two-loop order.

This calculation is valid for the higher $l$ modes as well, so in principle we could fit the $l=\{4,6$,$8\}$ modes using the same ten free parameters, in analogy to the calculation done for dark matter in \cite{Lewandowski:2015ziq}. However, the higher-$l$ modes are difficult to measure in simulations due to their small magnitude, and they were not available for this analysis. All in all, we find that the EFT gives a good fit to the simulated real-space halo power spectrum and the $l=0$ and $l=2$ modes of the redshift-space halo power spectrum at $z=0.67$ up to $k=0.43 \  \unitsk$. Though extremely good, the actual $k$-reach of the fit should be taken with care because, as noted for example in~\cite{Foreman:2015lca}, it is possible that the reach of the theory is somewhat overestimated when using just the one-loop expressions or not extremely accurate data. Using for example more accurate data or the two-loop expressions, which grow steeper at higher wavenumber, would allow a safer estimate of the $k$-reach. We plan to do this in future work.

\subsection{Fits to Galaxies\label{app:galaxies}}
In this subsection we show that that we can also fit to a comparable level of accuracy the effective theory to the power spectrum for a realistic model of galaxies in real space and redshift space~\footnote{More precisely, at the highest wavenumbers where we fit, the errors for the real-space dark matter, the real-space biased tracers, and the biased tracers monopole power spectra are less than 2\%. Instead, the error for the biased tracers power spectrum quadrupole is about 7\% for the haloes and 15\% for the $v_{m_{peak}}$ model of LRGs.}. This capability is indeed expected from the EFTofLSS point of view, because all biased tracers are equal at a conceptual level, and they differ only for the size of the bias parameters (see~\cite{Fujita:2016dne} for a discussion on how the $k$-reach is affected by different halo populations and how this might require the addition of higher order terms in order to reach the same accuracy at a given wavenumber). The fit to the power spectra of the $v_{m_{peak}}$ model of LRGs from \cite{Jennings:2015lea} is given in Fig.~\ref{vmpeak}. We find that the theory agrees with the data to within a few percent up to $k \sim 0.43 \ \unitsk$ (notice though that the error bars for $P_2$ are about 15\% in the relevant region.). This fit has the same reach of the theory as the fit to halos given in the main text, further demonstrating the consistency of the EFT. Note that what looks like a failure of the fit around $k \sim 0.34 \ \unitsk$ comes from the fact that the data for $P_2$ crosses zero there, so the ratio we are plotting diverges. This is just due to the choice of plotting the ratio of the two curves rather than the two curves directly, and it is not a failure of the theory. As we did for the halos, we perform a consistency check of our fitting procedure in Appendix \ref{check} by implementing a fitting procedure incorporating the estimated theoretical error, and find consistent results.

\begin{figure}[htb] 
\centering
 \includegraphics[width=8.2cm]{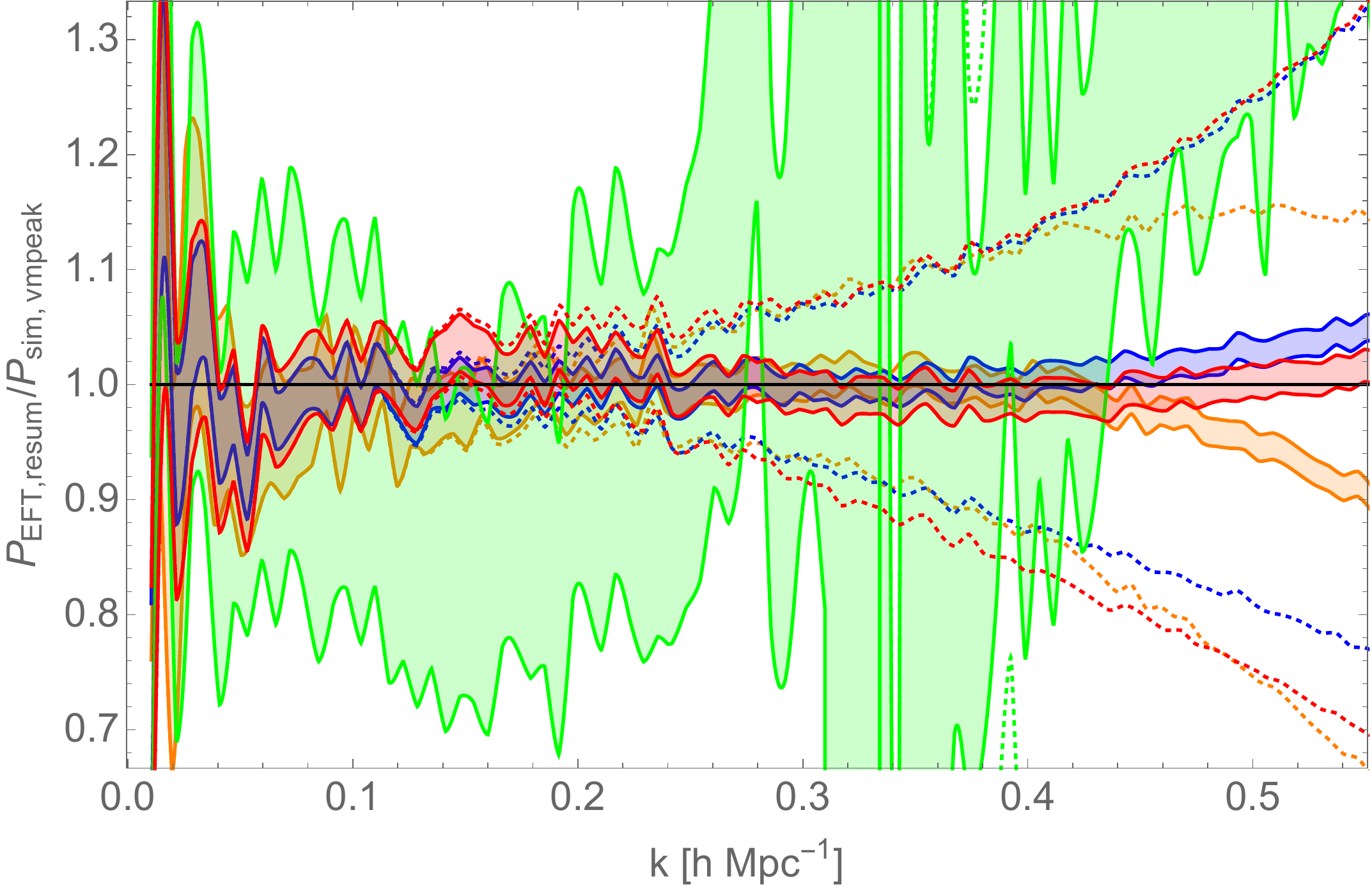}
 \includegraphics[width=8.2cm]{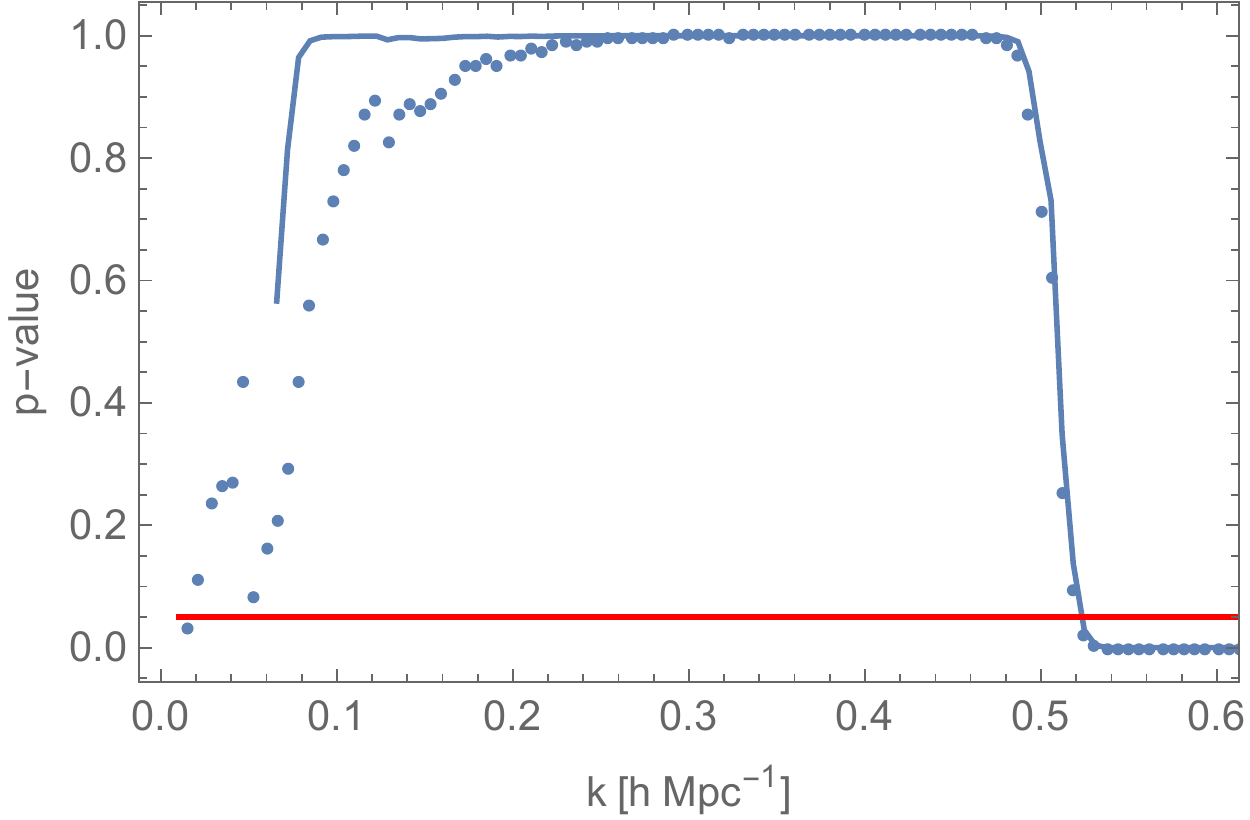}
  \caption{ {\it Left:} Results of the fits of the EFT power spectra at $z=0.67$ after IR-resummation to the power spectra of LRGs in the $v_{m_{peak}}$ sample~\cite{Jennings:2015lea}, which has a number density $\bar n=3.9 \cdot 10^{-4}(\hinvMpc)^3$, and dark matter extracted from simulations. The fits were performed  in the $k$-range $k_{\rm min}=0.01 \ \unitsk$ to $k_{\rm fit}=0.42 \ \unitsk$ and resulted in the best-fit parameters $\{ b_1=1.86 \pm 0.04,\ b_2=0.99\pm 7.59, \ b_3=-4.5 \pm 2.8, \ b_4=0.68 \pm 6.01,\ c_{\rm ct}^{(\delta_h)}=(0.69 \pm 8.35) \left( k_{\rm NL}  \ h^{-1} \ {\rm Mpc} \right)^{2}, \ \tilde{c}_{r,1}=(-30 \pm 6)\left( \km  \ h^{-1} \ {\rm Mpc} \right)^{2}, \ \tilde{c}_{r,2}=(4.6 \pm 1.3)\left( \km  \ h^{-1} \ {\rm Mpc} \right)^{2}, \ c_{\epsilon,1}=13 \pm 33, \ c_{\epsilon,2}=(30 \pm 12)\left( \km  \ h^{-1} \ {\rm Mpc} \right)^{2},\ c_{\epsilon,3}=(14 \pm 25)\left( \km  \ h^{-1} \ {\rm Mpc} \right)^{2} \}$ for the LRGs and $c_s^2=(-0.61 \pm 0.02) \left( k_{\rm NL}  \ h^{-1} \ {\rm Mpc} \right)^{2} $ for the dark matter. $P_{\rm real}$ is plotted in red, $P_{l=0}$ in blue, $P_{l=2}$ in green, and $P_{\rm DM}$ in orange. The shaded region shows the 1$\sigma$ error on the simulation data, which includes the error on the spectra from simulations described in \cite{Feldman:1993ky} and a $1\%$ error added in quadrature to account for unknown systematics. The expected theoretical error is given by the dotted lines. {\it Right:} Plot of $p$-values calculated up to a given $k$ for the IR-resummed fit to the $v_{m_{peak}}$ power spectra with $k_{\rm fit}=0.42 \ \unitsk$.  The solid blue curve shows the $p$-value, neglecting the data points with $k<0.06 \ {\rm h \ Mpc^{-1}}$, and the dotted blue curve includes all of the low-$k$ points. The horizontal red line shows $p=0.05$. }
     \label{vmpeak}
   \end{figure}

\section{Conclusion}
In this paper, we extend the work done in the EFTofLSS to derive the power spectrum of dark matter in redshift space \cite{Senatore:2014vja,Lewandowski:2015ziq}, and the power spectrum of halos in real space \cite{Senatore:2014eva,Angulo:2015eqa,Fujita:2016dne}, in order to calculate the power spectrum of halos in redshift space. We find that the power spectrum at one loop is given in terms of ten parameters. The four bias parameters, as well as the first ``speed of sound" counter-term and stochastic counter-term have already been seen in the power spectrum of halos in real space, and two of the redshift-space counter-terms appear already in the EFT of dark matter in redshift space. In addition to these we find  two novel higher-derivative stochastic bias parameters that need to be included for the full study of halos in redshift space. We see that the theory of biased tracers is extended to redshift space fairly easily, using only a few new integration kernels, because we can treat the velocity divergence as a special species of halos. In addition, we perform the IR-resummation of the halo power spectrum in redshift space, which follows directly from \cite{Lewandowski:2014rca}. In comparing to two populations of biased tracers, respectively with number density $\bar n=3.8 \cdot 10^{-2}(\hinvMpc)^3$ and $\bar n=3.9 \cdot 10^{-4}(\hinvMpc)^3$, which are measured from simulations, we find that the EFT agrees with the first two multipoles of the redshift-space halo power spectra and the real-space matter and halo power spectra at $z=0.67$ to within a few percent accuracy up to $k\simeq 0.43 \ \unitsk$~\footnote{More precisely, at the highest wavenumbers where we fit, the errors for the real-space dark matter, the real-space tracers, and the tracers monopole power spectra are less than 2\%, while  the error for the tracers power spectrum quadrupole is about 7\% for the haloes and 15\% for the $v_{m_{peak}}$ model of LRGs.}. We comment in the text on the fact that this actual value of the $k$-reach should be taken with care, and that, as pointed out in~\cite{Fujita:2016dne}, it will depend at some level on the galaxy population and the precision of the data. We also have excellent agreement with the BAO because we do not have large residual oscillations in the IR-resummed power spectra. 

In future work, we hope to be able to compare to higher multipoles, bispectra and to do so at additional redshifts. This will allow us to test and understand more in detail how sample size affects our results and to further check the consistency of the EFT approach, and in particular its $k$-reach.  We also plan to study how much each bias coefficient contributes to the fit, and, possibly, in this way to reduce the number of parameters to be measured from simulations (or observations). Finally, having the theory of biased tracers in redshift space means  we can explore the cosmological parameter constraints which can be achieved using the EFTofLSS framework on a realistic galaxy sample  at different redshifts, or, even better, directly to observations. We also hope to do this in future work.

%
%
 %
 %

\subsubsection*{Acknowledgments}
A.P. is partially supported by the Stanford Graduate Fellowship. L.S. is partially supported by DOE Early Career Award DE-FG02-12ER41854. E.J. is supported by Fermi Research Alliance, LLC under the U.S. Department of Energy under contract No. DEAC02-07CH11359.  This work received partial support from the U.S.\ Department of Energy under contract number DE-AC02-76SF00515.  This research made use of data from one of the Dark Sky Simulations, which were produced using an INCITE 2014 allocation (M. Warren et al.) on the Oak Ridge Leadership Computing Facility at Oak Ridge National Laboratory. We thank Sam Skillman, Mike Warren, Matt Turk and the Dark Sky collaboration for making these data available.

%
%
 %
 %

\begin{appendix}

\section{Halo kernels}\label{appendixa}

The expressions for the basis operators $\hat{c}^{(n)}_i$ were found in \cite{Angulo:2015eqa} and corrected in \cite{Fujita:2016dne}. We agree with the corrections, and they are reproduced below.
At first order, 
\be
\hat{c}_{\delta,1}^{(1)}=1 \ .
\ee 
At second order we have
\bea
\hat{c}_{\delta,1}^{(2)} &=& \frac{\q_1 \cdot \q_2}{q_1^2} \n
\hat{c}_{\delta,2}^{(2)} &=& F^{(2)}(\q_1,\q_2)-\frac{\q_1 \cdot \q_2}{q_1^2} \n
\hat{c}_{\delta^2,1}^{(2)} &=& 1 \n
\hat{c}_{s^2,1}^{(2)} &=& \frac{(\q_1 \cdot \q_2)^2}{q_1^2q_2^2}-\frac{1}{3}  \  , \n
\label{app:second}
\eea
and at third order,
\bea
\hat{c}_{\delta,1}^{(3)} &=&\frac{1}{2}\left( \frac{(\q_1 \cdot \q_2+\q_1 \cdot \q_3)}{(\q_2+\q_3)^2} G^{(2)}(\q_2,q_3)+\frac{(\q_1 \cdot \q_2)(\q_1 \cdot \q_3+\q_2 \cdot \q_3)}{q_2^2 q_3^2} \right) \n
\hat{c}_{\delta,2}^{(3)} &=& \frac{(\q_1 \cdot \q_3+\q_2 \cdot \q_3)}{q_2^2 q_3^2} \left( F^{(2)}(\q_1,\q_2)q_2^2-\q_1 \cdot \q_2 \right) \n
\hat{c}_{\delta,3}^{(3)} &=& F^{(3)}(\q_1,\q_2,\q_3)+\frac{(\q_1+\q_2)\cdot \q_3}{2 q_2^2 q_3^2}(\q_1\cdot \q_2-2 F^{(2)}(\q_1,\q_2)q_2^2) -\frac{\q_1 \cdot (\q_2+\q_3)}{2(\q_2+\q_3)^2}G^{(2)}(\q_2,\q_3) \n
\hat{c}_{\delta^2,1}^{(3)}&=& 2 \frac{\q_2 \cdot \q_3}{q_3^2} \n
\hat{c}_{\delta^2,2}^{(3)}&=& 2 F^{(2)}(\q_1,\q_2) -2 \frac{\q_2 \cdot \q_3}{q_3^2} \n
\hat{c}_{\delta^3,1}^{(3)}&=& 1 \n
\hat{c}_{s^2,1}^{(3)}&=& 2 \frac{\q_2 \cdot \q_3}{q_3^2}\left( \frac{(\q_1 \cdot \q_2)^2}{q_1^2 q_2^2} -\frac{1}{3}\right) \n
\hat{c}_{s^2,2}^{(3)}&=& 2 F^{(2)}(\q_1,\q_2) \left( \frac{((\q_1+ \q_2)\cdot \q_3)^2}{(\q_1+\q_2)^2 q_3^2} -\frac{1}{3}\right) - 2 \frac{(\q_2 \cdot \q_3)^2}{q_3^2}\left( \frac{(\q_1 \cdot \q_2)^2}{q_1^2 q_2^2} -\frac{1}{3}\right) \n
\hat{c}_{s^3,1}^{(3)}&=& (9 q_1^2 q_2^2 q_3^2)^{-1} \left( 9(\q_1 \cdot \q_2) (\q_1 \cdot \q_3) (\q_2 \cdot \q_3)-3(\q_1 \cdot \q_3) ^2 q_2^2-3(\q_1 \cdot \q_2) q_3^2-3(\q_2 \cdot \q_3)q_1^2+2 q_1^2q_2^2q_3^2  \right) \n
\hat{c}_{st}^{(3)}&=& \left(G^{(2)}(\q_1,\q_2)-F^{(2)}(\q_1,\q_2)  \right) \left( \frac{((\q_1+ \q_2)\cdot \q_3)^2}{(\q_1+\q_2)^2 q_3^2} -\frac{1}{3}\right) \n
\hat{c}_{\psi}^{(3)}&=& G^{(3)}(\q_1,\q_2,\q_3)-F^{(3)}(\q_1,\q_2,\q_3)+2 F^{(2)}(\q_1,\q_2) \left( F^{(2)}(\q_1+\q_2,\q_3)-G^{(2)}(\q_1+\q_2,\q_3) \right) \n
\hat{c}_{\delta s^2}^{(3)}&=&\frac{(\q_1 \cdot \q_2)^2}{q_1^2q_2^2}-\frac{1}{3}  \ . \n
\label{app:third}
\eea

From \cite{Angulo:2015eqa,Fujita:2016dne}, the coefficients of the new basis $\tilde{c}^{(A)}_i$ are related to the original coefficients $c^{(A)}_i$ as 
\bea
\tilde{c}^{(A)}_{\delta,1}&=&c^{(A)}_{\delta,1} \n
\tilde{c}^{(A)}_{\delta,2(2)}&=&\frac{7}{2}c^{(A)}_{s^2,1}+c^{(A)}_{\delta,2} \n
\tilde{c}^{(A)}_{\delta,2(3)}&=&\frac{7}{2}c^{(A)}_{s^2,1}+c^{(A)}_{\delta,2} \n
\tilde{c}^{(A)}_{\delta,3}&=&\frac{9}{2}c^{(A)}_{st,1}+\frac{45}{4}c^{(A)}_{s^3,1}+c^{(A)}_{\delta,3}+2 c^{(A)}_{\psi,1} \n
\tilde{c}^{(A)}_{\delta^2,1(2)}&=&-\frac{17}{6}c^{(A)}_{s^2,1}+c^{(A)}_{\delta^2,1} \n
\tilde{c}^{(A)}_{\delta^2,1(3)}&=&-\frac{17}{6}c^{(A)}_{s^2,1}+c^{(A)}_{\delta^2,1} \n
\tilde{c}^{(A)}_{\delta^2,2}&=&-\frac{71}{24}c^{(A)}_{st,1}-\frac{137}{16}c^{(A)}_{s^3,1}+c^{(A)}_{\delta^2,2}+\frac{7}{4} c^{(A)}_{\delta s^2,1}-\frac{55}{42}c^{(A)}_{\psi,1}\n
\tilde{c}^{(A)}_{s^2,2}&=&-\frac{1}{2}c^{(A)}_{st,1}+c^{(A)}_{s^2,2}-\frac{3}{4}c^{(A)}_{s^3,1}-\frac{2}{7}c^{(A)}_{\psi,1} \n
\tilde{c}^{(A)}_{\delta^3,1}&=&-\frac{17}{6}c^{(A)}_{s^2,1}+c^{(A)}_{\delta^2,1}  \ .
\eea

The choice of bias coefficients that make $\theta_h\equiv \delta_{\theta_h}$  are:
$\{\tilde{c}^{\theta_h}_{\delta,1}=1 ,
\tilde{c}^{\theta_h}_{\delta,2(2)}=2 ,
\tilde{c}^{\theta_h}_{\delta,2(3)}=2 ,
\tilde{c}^{\theta_h}_{\delta,3}=3 ,
\tilde{c}^{\theta_h}_{\delta^2,1(2)}=-1 ,
\tilde{c}^{\theta_h}_{\delta^2,1(3)}=-1 ,
\tilde{c}^{\theta_h}_{\delta^2,2}=-\frac{3}{2} ,
\tilde{c}^{\theta_h}_{s^2,2}=0 ,
\tilde{c}^{\theta_h}_{\delta^3,1}=1 \}$.

The power spectrum will be computed using the symmetrized version of these kernels with the UV part subtracted from the $\hat{c}^{(3)}_{i}$, so the relevant kernels are $\hat{c}^{(2)}_{i}(\kv-\qv,\qv)_{\sym}$ and $\hat{c}^{(3)}_{i}(\kv,-\qv,\qv)_ {\rm UV-sub,\sym}$:
\bea
\hat{c}^{(2)}_{\delta,1}(k,q,x)_{\sym} & =& \frac{-2 q^3+k^3 x+4 k q^2 x-k^2 q-2 k^2 q x}{2q(k^2+q^2-2 k q x)} +1\n
\hat{c}^{(2)}_{\delta,2}(k,q,x)_{\sym} &=& \frac{7 q^2-14 k q x+5 k^2+2 k^2 x^2}{7(k^2+q^2-2 k q x)} -1\n
\hat{c}^{(2)}_{\delta^2,1}(k,q,x)_{\sym} &=& 0  \ ,
\eea
and
\bea
&&\hat{c}^{(3)}_{\delta,1}(k,q,x)_{\rm UV-sub, \sym}  =\frac{13}{63} \frac{-7 k^6 x^2+28 k^4 q^2 x^2(x^2-1)-2 q^6(3+4 x^2)+k^2 q^4(44 x^4-17 x^2-6)}{42 q^2(k^2+q^2-2 k q x)(k^2+q^2-2 k q x)} \n
&&\hat{c}^{(3)}_{\delta,2}(k,q,x)_{\rm UV-sub, \sym}  = -\frac{4}{63}(3 x^2-1) \n
&&\hat{c}^{(3)}_{\delta,3}(k,q,x)_{\rm UV-sub, \sym}  = -\frac{4}{63}\frac{2 q^4(1-3 x^2+k^4(3 -8 x^2+x^4+k^2 q^2(5-22 x^2+25 x^4)))}{(k^2+q^2+2 k q x)(k^2+q^2-2 k q x)}\n
&&\hat{c}^{(3)}_{\delta^2,1}(k,q,x)_{\rm UV-sub, \sym} =0 \n
&&\hat{c}^{(3)}_{\delta^2,2}(k,q,x)_{\rm UV-sub, \sym}  = \frac{8}{63}(3 x^2-1) \n
&&\hat{c}^{(3)}_{\delta^3,1}(k,q,x)_{\rm UV-sub, \sym}  = 0 \n
&&\hat{c}^{(3)}_{s^2,2}(k,q,x)_{\rm UV-sub, \sym}  =\frac{58 q^4 (3 x^2-1)-k^4(119-267 x^2+90x^4)-2 k^2 q^2(74-235 x^2+219x^4)}{189(k^2+q^2+2 k q x)(k^2+q^2-2 k q x)} \ , \n
\label{irkern}
\eea
where $x=\frac{\kv \cdot \qv}{kq}$.

\section{Degeneracy of halo bias parameters}\label{sec3}
We now turn to the explicit calculation of the halo kernels in real space. In the ``basis of descendants" of \cite{Angulo:2015eqa}, which chooses the basis of linearly independent biases which gives priority to the various operators that descend from a given one by the Taylor expansion of $\xv_{\rm fl}$, the density of a general halo species ${A}$ is given as
\be
\delta^{(n)}_{A}(\kv)=\int d^3 q_1 \ldots d^3 q_n   \ K^{(n)}_{A}(\q_1, \ldots , \q_n)_{\sym} \delta^3_D(\kv-\qv_1 \ldots - \qv_n) \delta^{(1)}(\qv_1) \ldots \delta^{(1)}(\qv_n) \ ,
\ee
where these $K_{A,{\rm sym}}^{(n)}$ are the symmetrized versions of the following kernels:
\bea
K^{(1)}_{A}(\q_1) &=& \tilde{c}^{(A)}_{\delta,1}\hat{c}^{(1)}_{\delta,1}(\q_1) =\tilde{c}^{(A)}_{\delta,1}\n
K^{(2)}_{A}(\q_1,\q_2) &=&  \tilde{c}^{(A)}_{\delta,1}\hat{c}^{(2)}_{\delta,1}(\q_1,\q_2)+\tilde{c}^{(A)}_{\delta,2}\hat{c}^{(2)}_{\delta,2}(\q_1,\q_2) +\tilde{c}^{(A)}_{\delta^2,1}\hat{c}^{(2)}_{\delta^2,1}(\q_1,\q_2)\n
K^{(3)}_{A}(\q_1,\q_2,\q_3) &=& \tilde{c}^{(A)}_{\delta,1}\hat{c}^{(3)}_{\delta,1}(\q_1,\q_2,\q_3) + \tilde{c}^{(A)}_{\delta,2(3)}\hat{c}^{(3)}_{\delta,2}(\q_1,\q_2,\q_3) + \tilde{c}^{(A)}_{\delta,3}\hat{c}^{(3)}_{\delta,3}(\q_1,\q_2,\q_3) \n
&&+\tilde{c}^{(A)}_{\delta^2,1(3)}\hat{c}^{(3)}_{\delta^2,1}(\q_1,\q_2,\q_3)+\tilde{c}^{(A)}_{\delta^2,2}\hat{c}^{(3)}_{\delta^2,2}(\q_1,\q_2,\q_3) + \tilde{c}^{(A)}_{\delta^3,1}\hat{c}^{(3)}_{\delta^3,1}(\q_1,\q_2,\q_3) \n
&& +\tilde{c}^{(A)}_{s^2,2}\hat{c}^{(3)}_{s^2,2}(\q_1,\q_2,\q_3) \ ,
\label{hkernels}
\eea
and where the $\hat{c}^{(n)}_i$ are the eight independent bias kernels given in \eqn{app:second} and \eqn{app:third}. In terms of these kernels, the power spectrum of halos in real space is
\bea
\langle \delta_{A}(\kv) \delta_{A}(\kv) \rangle &=& \langle \delta_{A}^{(1)}(\kv) \delta_{A}^{(1)}(\kv) \rangle+ \langle \delta_{A}^{(2)}(\kv) \delta_{A}^{(2)}(\kv) \rangle+ 2 \langle \delta_{A}^{(1)}(\kv) \delta_{A}^{(3)}(\kv) \rangle \n
&& +\langle \delta_{A}(\kv) \delta_{A}(\kv) \rangle_{ct}+\langle \delta_{A}(\kv) \delta_{A}(\kv) \rangle_{\epsilon}\n
&=& (K^{(1)}_{\rm A})^2P_{11}(k)+2 \int d^3 \q \   \left(K^{(2)}_{\rm A}(\q,\kv-\q)_{\rm sym}\right)^2 P_{11}(|\kv-\q|)P_{11}(q) \n
&&+6\int d^3 \q  \ K^{(3)}_{\rm A}(\q,-\q, \kv)_{\rm sym} K^{(1)}_{\rm A}P_{11}(q) P_{11}(k) + \langle \delta_A(\kv) \delta_A(\kv) \rangle_{ct} \n
&&+ \langle \delta_{A}(\kv) \delta_{A}(\kv) \rangle_{\epsilon} \ .
\label{finalps}
\eea
Explicitly, the symmetrized second order kernel for the halo density is:
\bea
K^{(2)}_{\delta_h} (k,q,x)_{\sym} &=&  \frac{ \tilde{c}_{\delta,1}}{2q}\frac{-2 q^3+k^3 x+4 k q^2 x-k^2q(1+2 x^2)}{k^2+q^2-2 k q x} \n
&&+ \frac{\tilde{c}_{\delta,2(2)}}{7}\frac{7 q^2-14 k q x+k^2(5+2 x^2)}{k^2+q^2-2 k q x}+\tilde{c}_{\delta^2,1(2)}  \ ,
\label{k2exp}
\eea
which contains the three bias coefficients from the unsymmetrized kernel in \eqn{hkernels}. 

Let us now turn to the more complicated third-order kernel. In the calculation, instead of using $K_A^{(3)}$, we will actually use the UV-subtracted third-order kernel to make the integrals converge better. This UV-subtraction is defined as 
\be
K_{A}^{(3)}(k,q,x)_{\rm UV-sub}=K_A^{(3)}(k,q,x) - \lim_{\frac{q}{k} \to \infty} K_A^{(3)}(k,q,x) \ ,
\ee
where the explicit expressions of the UV-subtracted $\hat{c}_i^{(3)}$ are given in \eqn{irkern}. We are free to do this because this change will be absorbed in a change of the counter-terms. Notice that in the $q$ integral of the third line of \eqn{finalps}, $K_A^{(3)}$ is the only term that has dependence on the angular coordinate $x$, so we are free to perform the $x$ integral on the kernel itself. After doing this integral we find that the final third-order kernel is
\bea
&&K^{(3)}_{\delta_h} (k,q)_{\rm UV-sub,\sym} =  \frac{ \tilde{c}_{\delta,1} }{504 k^3 q^3} \left(-38 k^5 q +48 k^3q^3-18 k q^5+9(k^2-q^2)^3\log \left[\frac{k-q}{k+q} \right] \right)  \n
&& \qquad \qquad +\frac{\tilde{c}_{\delta,3}+15\tilde{c}_{s^2,2} }{756 k^3 q^5} \left(2 k q (k^2+q^2)(3k^4-14 k^2 q^2+3 q^4)+3(k^2-q^2)^4\log\left[\frac{k-q}{k+q} \right] \right)  \ . \n
\label{hkernelsfin}
\eea
We see that after integration, $K_A^{(3)}$ only contains three bias parameters, rather than the seven it had in \eqn{hkernels}. This is because the momentum kernels multiplying the other four bias parameters have integrated to zero. Of the three that remain, $\tilde{c}_{\delta,1}$ has already appeared in the second order kernel, and $\tilde{c}_{\delta,3}$ and $\tilde{c}_{s^2,2}$ can be combined into one independent parameter. Thus we can define the following four independent bias parameters at one-loop \cite{Angulo:2015eqa}:
\bea
b_{1} &=& \tilde{c}_{\delta,1} \n
b_{2}&=&\tilde{c}_{\delta,2(2)} \n
b_{3}&=&\tilde{c}_{\delta,3}+15\tilde{c}_{s^2,2} \n
b_{4}&=&\tilde{c}_{\delta^2,1(2)} \ .
\eea
It was only after the angular integral that the degeneracies in the $\hat{c}_i^{(n)}$ operators became fully apparent. At higher loops, $K^{(3)}_A$ will contract with kernels that have nontrivial angular dependence, so the cancellations we encountered in \eqn{hkernelsfin} will not occur and the other four bias parameters will become important. However, at one loop, the halo density power spectrum is fully described by these four bias parameters plus the biased dark matter counter-term parameter, which we will discuss in the main text along with the counter-terms from the transformation to redshift space.

Finally, the velocity divergence power spectrum is described by the following kernels 
\bea
K^{(1)}_{\theta_h} (k,q,x)_{\sym} &=& 1  \n
K^{(2)}_{\theta_h} (k,q,x) _{\sym}&=& \frac{k^2(7 k x -q(1+6 x^2))}{14 q(k^2+q^2-2 k q x)} \n
K^{(3)}_{\theta_h} (k,q,x)_{\rm UV-sub,\sym}&=&  \frac{12 k^7 q -82 k^5 q^3+4 k^3 q^5-6 k q^7+3(k^2-q^2)^3(2k^2+q^2)\log \left[ \frac{k-q}{k+q}\right]}{504 k^3 q^5} \ ,  \n
\label{hkernelsfintheta}
\eea
with no additional free parameters. Now that we have the explicit expressions for the halo density and velocity power spectra in real space, all that remains is to transform to redshift space.

\section{Redshift-space kernels}\label{rskernels}
The contact terms in the redshift-space expansion generate new momentum kernels. The new terms coming from the terms in brackets in \eqn{brackets} at second order are
\bea
\delta_{[\frac{\partial_z}{\partial^2} \theta_h \delh]}^{(2)} (\kv) &=&\int_{\q_1, \q_2} \left( \frac{-i q_{1z}} {q_1^2} \right)  \theta_h^{(1)}(q_1)\delh^{(1)}(q_2) \delta^3_D(\kv -\q_1-\q_2) \n
&=& \int_{\q_1, \q_2} \left( \frac{-i q_{1z}} {q_1^2} \right)  K_{\theta_h} ^{(1)}K_{\delta_h} ^{(1)}\delta^{(1)}(q_1)\delta^{(1)}(q_2) \delta^3_D(\kv -\q_1-\q_2) \n
\delta_{[\frac{\partial_z}{\partial^2} \theta_h \frac{\partial_z}{\partial^2} \theta_h]}^{(2)} (\kv) &=&\int_{\q_1, \q_2}  \left( \frac{-i q_{1z}} {q_1^2} \right) \left( \frac{-i q_{2z}} {q_2^2} \right)  \theta_h^{(1)}(q_1)\theta_h^{(1)}(q_2) \delta^3_D(\kv -\q_1-\q_2) \n
&=& \int_{\q_1, \q_2} \left( -\frac{q_{1z}q_{2z}}{q_1^2 q_2^2} \right) K_{\theta_h} ^{(1)} K_{\theta_h} ^{(1)} \delta^{(1)}(q_1)\delta^{(1)}(q_2) \delta^3_D(\kv -\q_1-\q_2)  \ ,\n
\label{rsec}
\eea
and the new terms at third order are 
\bea
\delta_{[\frac{\partial_z}{\partial^2} \theta_h \frac{\partial_z}{\partial^2} \theta_h \frac{\partial_z}{\partial^2} \theta_h]}^{(3)} (\kv) &=& \int_{\q_1, \q_2 , \q_3}  \left( \frac{-i q_{1z}} {q_1^2} \right)  \left( \frac{-i q_{2z}} {q_2^2} \right)  \left( \frac{-i q_{3z}} {q_3^2} \right)  \theta_{h}^{(1)}(\q_1) \theta_{h}^{(1)}(\q_3) \theta_{h}^{(1)}(\q_3) \n
&& \qquad \qquad \delta^3_D(\kv -\q_1-\q_2-\q_3) \n
 &=& \int_{\q_1, \q_2 , \q_3}  \left(i \frac{ q_{1z}q_{2z}q_{3z}}{q_1^2 q_2^2 q_3^2}  \right)K_{\theta_h} ^{(1)} K_{\theta_h} ^{(1)}K_{\theta_h} ^{(1)} \delta^{(1)}(\q_1)\delta^{(1)}(\q_2)\delta^3_D(\kv -\q_1-\q_2-\q_3)  \n
\delta_{[\frac{\partial_z}{\partial^2} \theta_h \frac{\partial_z}{\partial^2} \theta_h  \delh]}^{(3)} (\kv) &=& \int_{\q_1, \q_2 , \q_3}   \left( \frac{-i q_{1z}} {q_1^2} \right)  \left( \frac{-i q_{2z}} {q_2^2} \right)    \theta_h^{(1)}(q_1)\theta_h^{(1)}(q_2) \delh^{(1)}(q_3) \delta^3_D(\kv -\q_1-\q_2-\q_3) \n
&=& \int_{\q_1, \q_2 , \q_3}   \left(-\frac{ q_{1z}q_{2z}}{q_1^2 q_2^2}  \right)K_{\theta_h} ^{(1)} K_{\theta_h} ^{(1)} K_{\delta_h} ^{(1)} \delta^{(1)}(\q_1)\delta^{(1)}(\q_2)  \delta^3_D(\kv -\q_1-\q_2 -\q_3) \n
\delta_{[\frac{\partial_z}{\partial^2} \theta_h \delh]}^{(3)} (\kv) &=& \int_{\q_1, \q_2} \left( -\frac{i q_{2z}}{q_2^2} \delh^{(2)}(q_1) \theta_h^{(1)}(q_2) - \frac{i q_{1z}}{q_1^2} \theta_h^{(2)}(q_1)\delh^{(1)}(q_2)  \right) \delta^3_D(\kv -\q_1-\q_2) \n
&=&   \int_{\q_1, \q_2, \q_3} \  \left( -\frac{i q_{3z}}{q_3^2}K_{\delta_h} ^{(2)}(\q_1, \q_2)  K_{\theta_h} ^{(1)}-  \frac{i (\q_1+\q_2)_z}{(\q_1+\q_2)^2}K_{\theta_h} ^{(2)}(\q_1, \q_2) K_{\delta_h} ^{(1)} \right)  \n
&& \qquad \qquad \delta^{(1)}(q_1) \delta^{(1)}(\q_2)\delta^{(1)}(\q_3) \delta^3_D(\kv-\q_1-\q_2-\q_3)  \n
\delta_{[\frac{\partial_z}{\partial^2} \theta_h \frac{\partial_z}{\partial^2} \theta_h]}^{(3)} (\kv) &=& \int_{\q_1, \q_2}\left( \frac{-i q_{1,z}}{q_1^2} \right) \left(\frac{-i q_{2,z}}{q_2^2} \right)2 \theta_h^{(2)}(q_1)\theta_h^{(1)}(q_2) \delta^3_D(\kv -\q_1-\q_2) \n
&=&  \int_{\q_1, \q_2, \q_3} \left( -\frac{(\q_1+\q_2)_z q_{3z}}{(\q_1+\q_2)^2 \q_3^2} \right)2K_{\theta_h} ^{(2)}(\q_1,\q_2)K_{\theta_h}^{(1)}\delta^{(1)}(\q_1) \delta^{(1)}(\q_2)  \delta^{(1)}(\q_3) \n
&& \qquad \qquad \delta^3_D(\kv -\q_1-\q_2-\q_3) \ . \n
\label{rthird}
\eea

\section{The IR-safe integrand}
Equal-time correlators in the EFT of LSS generically have IR divergences that cancel between diagrams after integration. This cancellation is difficult to implement precisely in numerical calculations, so it is useful to find a formulation of the integrand that is manifestly IR-safe. This was developed for dark matter correlators in \cite{Carrasco:2013sva}, and we will extend it to halos in real space and redshift space in this appendix.

For halos in real space, the extension is quite simple. It can be checked that all of the kernels in \eqn{app:second} and \eqn{app:third} are finite in the limit $q \to 0$ except for  $\hat{c}_{\delta,1}^{(2)}$ and $\hat{c}_{\delta,1}^{(3)}$. These kernels come into the power spectrum through the halo kernels $K^{(2)}_{A}$ and $K^{(3)}_{\rm A}$ as:
\bea
P_{AB}^{(22)}(k) &=& 2 \int d^3 \q \   K^{(2)}_{A}(\q,\kv-\q)_{\rm sym}K^{(2)}_{B}(\q,\kv-\q)_{\rm sym} P_{11}(|\kv-\q|)P_{11}(q) \n
P_{AB}^{(13)}(k) &=& 3\int d^3 \q  \ \left(K^{(3)}_{A}(\q,-\q, \kv)_{\rm sym} K^{(1)}_{B} +K^{(3)}_{B}(\q,-\q, \kv)_{\rm sym} K^{(1)}_{A} \right)P_{11}(q) P_{11}(k)  \ .
\label{p22p13}
\eea
Since $K_{A}^{(2)}$ comes into the power spectrum multiplied by $K_{B}^{(2)}$, there are also sub-leading divergences as $q \to 0$ that are generated when the non-divergent kernels multiply $\hat{c}_{\delta,1}^{(2)}$. From the explicit expressions for the halo kernels given in \eqn{k2exp} and \eqn{hkernelsfin}, we find that the IR-divergent contributions to $P_{AB}^{(22)}$ are:
\bea
P_{AB,{\rm IR}}^{(22)}(k) &=& \int_{q\ll k}  d^3q \   P_{11}(q)P_{11}(k)\left(\frac{ k^2 x^2 b_1^{(A)} b_1^{(B)} }{2 q^2}  - \frac{7b_1^{(A)}b_1^{(B)}k x}{q}  \right. \n
&&\left.  +\frac{b_1^{(A)} k x}{q} \left(7 b_4^{(B)}+b_2^{(B)}\left(5+2 x^2\right)\right)+ \frac{b_1^{(A)}k x}{q} \left(7 b_4^{(B)}+b_2^{(B)}\left(5+2 x^2\right)\right) \right) \ .
\label{sym22}
\eea
All of the terms proportional to $q^{-1}$ are odd in the angular variable $x$, so we can make them IR-safe by symmetrizing the integrand under the exchange $\qv \to -\qv$. In contrast, $P_{AB}^{(13)}(k)$ is automatically symmetric in $x$, and gives the following divergent term as $q\to 0$:
\bea
P_{AB,{\rm IR}}^{(13)}(k) &=& -\int_{q\ll k} d^3q \frac{ k^2 x^2  b_1^{(A)} b_1^{(B)}}{q^2}P_{11}(q)P_{11}(k) \ .
\label{auto13}
\eea

The leading IR divergences from the integrands in \eqn{sym22} and \eqn{auto13} do not  quite cancel. However, the integrand of $P_{AB}^{(22)}(k)$ has an additional divergence as $\kv \to \qv$ that we need to investigate. Since we used the symmetrized kernel in \eqn{p22p13}, the integrand is symmetric under the exchange $\q  \to \kv-\qv$. For any integrand $f(\kv,\qv)$ that has this symmetry, we can write the integral over $\qv$ as:
\bea
\int d^3 q \ f(\kv,\qv) &=&\int_{|\qv|<|\kv-\qv|} d^3q \  f(\kv,\qv)+\int_{|\qv|>|\kv-\qv|} d^3q \   f(\kv,\qv) \n
&=&\int_{|\qv|<|\kv-\qv|} d^3q  \ f(\kv,\qv)+\int_{|\vec{p}|<|\kv-\vec{p}|} d^3p \ f(\vec{k},\kv-\vec{p}) = 2\int d^3q \  f(\kv,\qv) \Theta(|\qv|-|\kv-\qv|) \ . \n
\label{trick}
\eea
Thus, we can map the $\kv \to \qv$ divergence of $P_{AB}^{(22)}$ into a $q \to 0$ divergence. Implementing \eqn{trick} and symmetrizing with respect to $\qv \to -\qv$  gives \bea
P_{AB,{\rm IR-safe}}^{(22)}(k) &=&\int d^3 q \  \biggl(  K^{(2)}_{A}(\q,\kv-\q) K^{(2)}_{B}(\q,\kv-\q)  P_{11}(|\kv-\qv |)P_{11}(q) \Theta(|\qv|-|\kv-\qv |) \n
&&  \qquad   + K^{(2)}_{A}(-\q,\kv+\q) K^{(2)}_{B}(-\q,\kv+\q)   P_{11}(|\kv+\qv |)P_{11}(q) \Theta(|\qv|-|\kv+\qv |) \biggr) \ . \n
\label{halosafe}
\eea
Now in the $\qv \to 0$ limit, the two terms in \eqn{halosafe} give a factor of two that makes the leading divergence of \eqn{sym22} cancel with \eqn{auto13}, and the sub-leading divergences from \eqn{sym22} are zero due to symmetrization. This makes the total integrand IR-safe.

When we go to redshift space, the new kernels $K_{h,r}^{(n)}$ contain the real-space halo kernels $K_{A}^{(n)}$, plus additional terms that are proportional to powers of $\mu$ and have new IR divergences. Let us first consider $P_{h,r}^{(13)}(k)$. We know that the only IR-divergent part of $K^{(3)}_A$  came from the kernel $\hat{c}^{(3)}_{\delta,1}$, so none of the bias parameters other than $b_1$ appear in the $q \to 0$ limit. 
Indeed, in the $q\to 0$ limit, we find that the divergent part of $P_{h,r}^{(13)}(k)$ is
\be
\ P_{h,r,{\rm IR}}^{(13)}(k)=-\int_{q \ll k} d^3 q \ \frac{ \pi k^2 (b_1+f \mu^2)^2(f^2 \mu^2(1-\mu^2)+x^2(2+f\mu^2(4+f(3\mu^2-1))))}{q^2}P_{11}(q)P_{11}(k) \ .
\ee

We turn now to $P_{h,r}^{(22)}(k)$. Just as for the halos in real space, the terms in $K_{h,r}^{(2)}$ proportional to $b_2$, $b_3$, and $b_4$ are finite. However, since $K_{h,r}^{(2)}$ comes into the power spectrum squared, we have divergent terms proportional to these bias parameters that come from the non-IR-divergent part of the halo kernel multiplying IR-divergent terms coming from the transformation to redshift space. In the limit $q \to 0$, all of these terms are odd in the angular variable $x$, so we can again symmetrize in $\qv \to -\qv$ to cancel them. After symmetrization, the $q\to 0$ limit of $P_{h,r}^{(22)}(k)$ is
\bea
P_{h,r,{\rm IR}}^{(22)}(k)&=&\int_{q \ll k} d^3 q \ \frac{ \pi k^2 (b_1+f \mu^2)^2(f^2 \mu^2(1-\mu^2)+x^2(2+f\mu^2(4+f(3\mu^2-1))))}{2q^2}P_{11}(q)P_{11}(k)  \n
&=&-\frac{1}{2}P_{h,r,IR}^{(13)}(k)  \ .
\label{ir22}
\eea

Since $K^{(2)}_{h,r}$ is symmetric in $\qv  \to \kv-\qv$, we can again use the trick in \eqn{trick} to rewrite the $P_{h,r,IR}^{(22)}(k)$ integral in a form that cancels exactly with $P^{(13)}_{h,r,IR}(k)$. Thus we find that the IR-safe integrand for halos in redshift space at one loop generalizes from the IR-safe integrand of dark matter in real space, and entails rewriting $\left<\delta_{h,r}\delta_{h,r}\right>_{22}+\left<\delta_{h,r}\delta_{h,r}\right>_{13}$ as follows:
\bea
\left<\delta_{h,r}\delta_{h,r}\right>_{22}+\left<\delta_{h,r}\delta_{h,r}\right>_{13}&=&\int d^3 q \  \biggl( 2 \left(K^{(2)}_{h,r}(\q,\kv-\q)_{\sym}\right)^2 P_{11}(|\kv-\qv |)P_{11}(q) \Theta(|\qv|-|\kv-\qv |) \n
&&  \qquad   +2 \left(K^{(2)}_{h,r}(-\q,\kv+\q)_{\sym}\right)^2 P_{11}(|\kv+\qv |)P_{11}(q) \Theta(|\qv|-|\kv+\qv |)  \n
&&   \qquad+  6 \left(K^{(3)}_{h,r}(\q,-\q, \kv)_{\sym} K^{(1)}_{h,r} \right) P_{11}(q) P_{11}(k) \biggl) \ .
\eea

\section{More details of the IR-resummation}\label{irResum}

In this appendix we give the explicit procedure for the IR-resummation discussed in Section \ref{secir}. The IR-resummation procedure in redshift space involves computing the factor $\tilde{K}_0$ defined in \eqn{k0def}, which is given explicitly by the expression
\be
\tilde{K}_0(\kv, \qv)=\exp\left[-\frac{k^2}{2}X_1(q)\left(1+2f \mu^2+f^2 \mu^2\right)-\frac{k^2}{2}Y_1(q)\left((\hat{k}\cdot\hat{q})^2 +2 f \mu(\hat{q}\cdot\hat{z}) (\hat{k}\cdot\hat{q})+f^2 \mu^2 (\hat{q}\cdot\hat{z})^2 \right)\right] \ , \n
\label{k0exp}
\ee
where $X_1(q)$ and $Y_1(q)$ are the following functions of the linear halo power spectrum:
\bea
X_1(q)&=&\frac{1}{2 \pi^2}\int_0^\infty dk \ e^{\left(-k^2/\Lambda_{\rm IR}\right)}P_{hh}(k) \left( \frac{2}{3}-2 \frac{j_1(kq)}{kq}\right) \n
Y_1(q)&=&\frac{1}{2 \pi^2}\int_0^\infty dk \ e^{\left(-k^2/\Lambda_{\rm IR}\right)}P_{hh}(k) \left(-2j_0(kq)+6 \frac{j_1(kq)}{kq}\right)  \ ,
\label{xeq}
\eea
and where the final answer is independent of the specific value of $\Lambda_{IR}$ as long as it includes all the relevant modes that need to be resummed.

The authors of \cite{Lewandowski:2015ziq} noticed that due to the infinite radius of convergence of the exponential function, a Taylor series expansion can be used even for a non-infinitesimal argument of the exponential as long as enough terms are kept to reach the desired precision. They found that for the one-loop dark matter power spectrum it was sufficient to expand \eqn{k0exp} to third order in $\mu^2 k^2 X_1(q)$ and to first order in $k^2 Y_1(q)$.

Using the same order of approximation as \cite{Lewandowski:2015ziq}, we find that the halo power spectra in redshift space can be resummed as follows:
\bea
P_{h, {\rm linear}}^{l, {\rm resum}} (k)&=&\sum_{l=0,2,4}\int d^3 q \left[ P_{h, {\rm linear}}^{l'}(q) M^{(1)}_{l,l'}(k)+\frac{2l+1}{2}e^{-\frac{k^2}{2}X_1(q)}\left(\left(1+\frac{k^2}{2}X_1(q) \right)I^0_{l,l'}(k,q) \right. \right. \n
&& \qquad \qquad \qquad \left. \left.+\frac{k^2}{2}f(2+f)X_1(q) I^2_{l,l'}(k,q)\right)P_{h, {\rm linear}}^{l'}(k) \right] \n
P_{h, {\rm 1-loop}}^{l,{\rm resum}} (k)&=&\sum_{l=0,2,4,6,8}\int d^3 q \left[ P_{h, {\rm 1-loop}}^{l'}(q) M^{(0)}_{l,l'}(k)+\frac{2l+1}{2}e^{-\frac{k^2}{2}X_1(q)}I^0_{l,l'}(k,q)P_{h, {\rm 1-loop}}^{l'}(k) \right] \ , \n
\label{irresum}
\eea
where $M^{(n)}_{l,l'}$ is given in \eqn{meq} and
\be
I_{l,l'}^a=\int_{-1}^1 d\mu  \ \mathcal{P}_l(\mu) \mathcal{P}_{l'}(\mu) \mu^a e^{-\frac{k^2}{2}X_1(q)\mu^2 f(2+f)} \ .
\label{ieq}
\ee

\begin{figure}[htb] 
\centering
 \includegraphics[width=9.5cm]{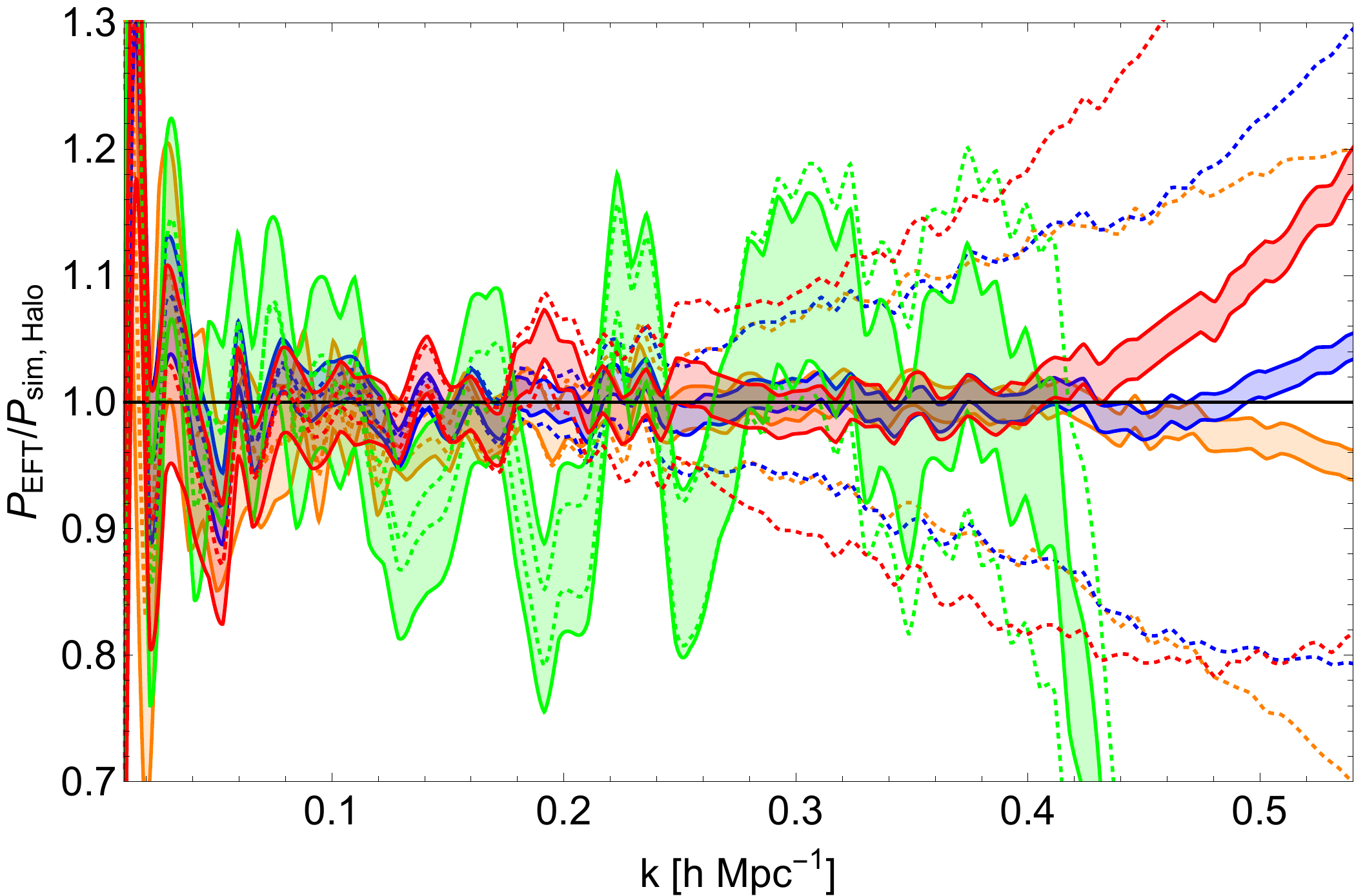}
  \caption{  Results of the fits of the EFT power spectra at $z=0.67$ before IR-resummation to the power spectra of halos and dark matter extracted from simulations, which were performed  in the $k$-range $k_{\rm min}=0.01 \ \unitsk$ to $k_{\rm fit}=0.39 \ \unitsk$ and resulted in the best-fit parameters $\{ b_1=0.98 \pm 0.01 ,\ b_2=1.4 \pm 1.9, \ b_3=-0.84 \pm 0.88, \ b_4=-0.83\pm 1.63,\ c_{\rm ct}^{(\delta_h)}=(9.6 \pm 3.0)\left( k_{\rm NL} \ h^{-1}  {\rm Mpc} \right)^{2}, \ \tilde{c}_{r,1}=(-12 \pm 4) \left(\km \ h^{-1}  {\rm Mpc} \right)^{2}, \ \tilde{c}_{r,2}=(-0.45 \pm 1.26)\left(\km \ h^{-1}  {\rm Mpc} \right)^{2}, \ c_{\epsilon,1}=-1.4 \pm 10.7, \ c_{\epsilon,2}=(11 \pm 2)\left(\km \ h^{-1}  {\rm Mpc} \right)^{2},\ c_{\epsilon,3}=(-7.1\pm 8.2)\left(\km \ h^{-1}  {\rm Mpc} \right)^{2}\}$ for the halos and $c_s^2=(-0.61 \pm  0.02) \left( k_{\rm NL} \ h^{-1}  {\rm Mpc} \right)^{2}$ for the dark matter.  $P_{\rm real}$ is plotted in red, $P_{l=0}$ in blue, $P_{l=2}$ in green, and $P_{\rm DM}$ in orange. The shaded region shows the 1$\sigma$ error on the simulation data, which includes the error on the halo spectra from simulations described in \cite{Feldman:1993ky} and a $1\%$ error added in quadrature to account for unknown systematics. The expected theoretical error is given by the dotted lines.  }
     \label{noresum}
   \end{figure}
   
The results of this resummation procedure are used for the fits in Fig.~\ref{fits}, while the non-IR-resummed fits are given in Fig.~\ref{noresum}. Comparing the two figures, we see that the IR resummation was necessary to achieve a good fit to the simulation data, especially for the $l=2$ mode.

\section{A Further Check of the Fitting Procedure}\label{check}
In this appendix, we implement a different fitting procedure that incorporates the estimated theoretical error in order to further test that we are not overfitting~\cite{Bertolini:2016hxg}. Since the theoretical error is estimated only at the order of magnitude level, the results of this section should be taken more as a reasonable consistency check rather than as an absolute check. In this procedure we use all of the data up to $k=0.54 \ \unitsk$, a point well past where the theory is expected to fail, and we include the theoretical error described in footnote \ref{theory}, added in quadrature to the data error and the systematic error used for the fits in Section \ref{secfits}. The theoretical error is added in order to account for the larger uncertainty at high $k$, so that these data can still be used in the fit without however biasing the results. The results of this fitting procedure are given along with a plot of the $p$-value in Fig.~\ref{newhalo} for the halos and Fig.~\ref{newLRG} for the $v_{m_{peak}}$ sample. We find that the parameters obtained in this new fitting procedure agree with the parameters obtained in Section \ref{secfits} to within 1$\sigma$ for the halos and 1.8$\sigma$  for the LRGs (in this case only two parameters are beyond 1$\sigma$, one at $1.2\sigma$ and the other at $1.8\sigma$). Futhermore, the $p$-value plots in Fig.~\ref{newhalo} and Fig.~\ref{newLRG} indicate that the fits perform well up to about the same $k$.

This procedure is somewhat uncertain, as the theory error is only known to within an order of magnitude. However, it is encouraging that the results of this procedure are roughly consistent with the fits performed in Section \ref{secfits}.

\begin{figure}[htb] 
\centering
 \includegraphics[width=8.2cm]{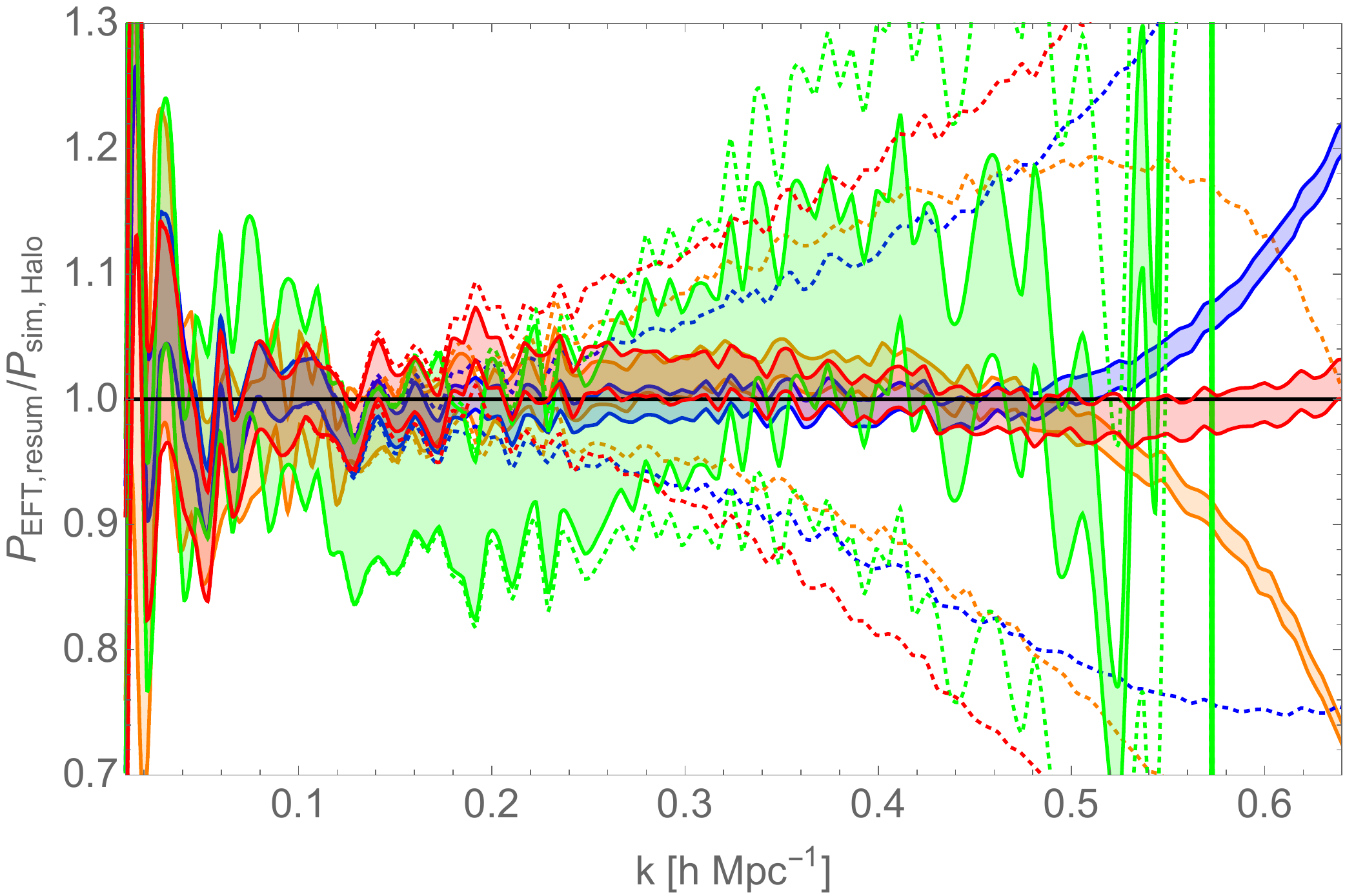}
 \includegraphics[width=8.2cm]{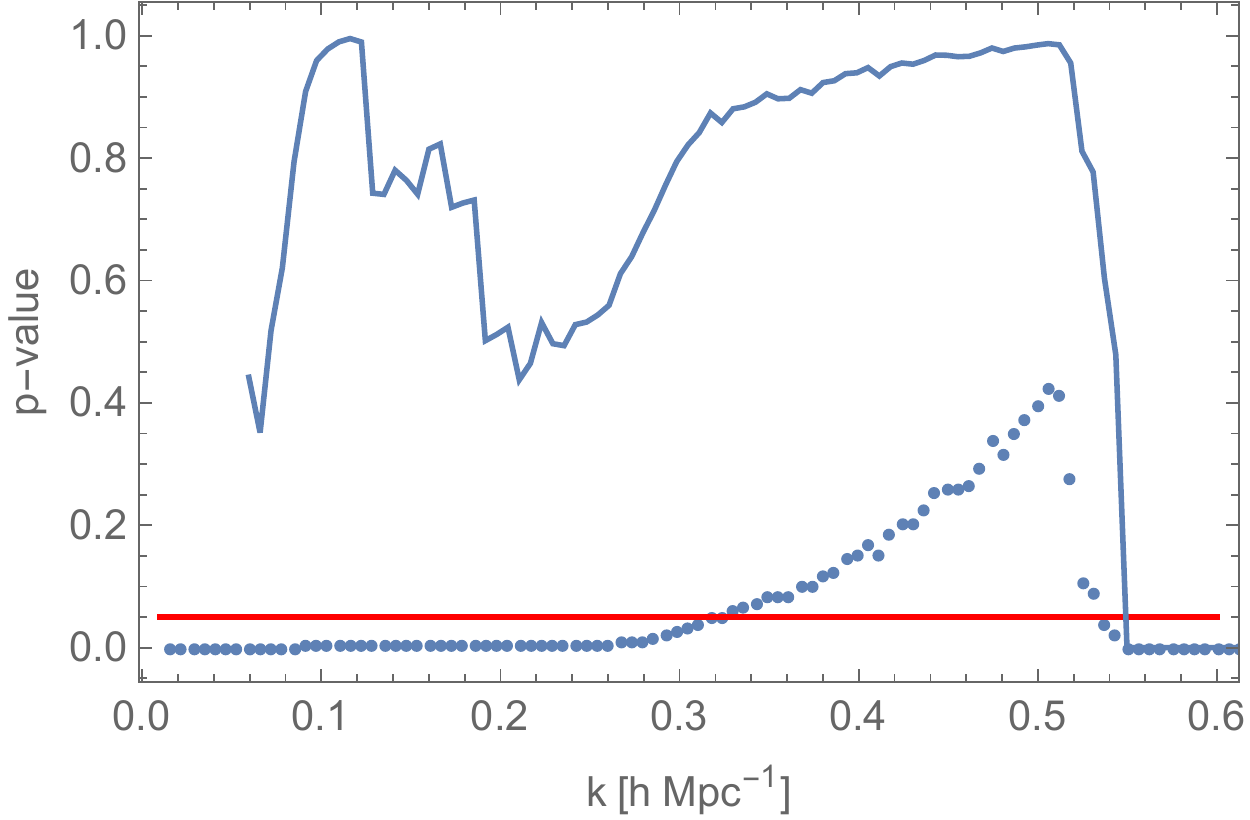}
  \caption{ {\it Left:} {Results of the fits, including theoretical error in quadrature, of the EFT power spectra at $z=0.67$ after IR-resummation to the power spectra of halos \cite{Jennings:2015lea}, which has a number density $\bar n=3.8 \cdot 10^{-2}(\hinvMpc)^3$, and dark matter extracted from simulations. The fits were performed  in the $k$-range $0.01 \ \unitsk$ to $0.54 \ \unitsk$ and resulted in the best-fit parameters $\{ b_1=0.98 \pm 0.01, b_2= 0.04 \pm 1.28, b_3 = 0.06 \pm 0.90,  b_4 = 0.56 \pm 1.05,  c_{\rm ct}^{(\delta_h)}= (6.2 \pm 3.2)\left( k_{\rm NL}  \ h^{-1} \ {\rm Mpc} \right)^{2}, \tilde{c}_{r,1} = (-15 \pm 2)\left( \km  \ h^{-1} \ {\rm Mpc} \right)^{2}, \tilde{c}_{r,2} =( 0.64 \pm 0.85)\left( \km  \ h^{-1} \ {\rm Mpc} \right)^{2}, c_{\epsilon,1} = 1.1 \pm 8.4,  c_{\epsilon,2} = (4.6 \pm 1.0)\left( \km  \ h^{-1} \ {\rm Mpc} \right)^{2},  c_{\epsilon,3} = (12 \pm 1 )\left( \km  \ h^{-1} \ {\rm Mpc} \right)^{2}\}$ for the halos and $c_s^2=(-0.49 \pm 0.01)\left( k_{\rm NL}  \ h^{-1} \ {\rm Mpc} \right)^{2} $ for the dark matter. $P_{\rm real}$ is plotted in red, $P_{l=0}$ in blue, $P_{l=2}$ in green, and $P_{\rm DM}$ in orange. The shaded region shows the 1$\sigma$ error on the simulation data, which includes the error on the spectra from simulations described in \cite{Feldman:1993ky} and a $1\%$ error added in quadrature to account for unknown systematics. The expected theoretical error is given by the dotted lines. {\it Right:} Plot of $p$-values calculated up to a given $k$ for the IR-resummed fit to the $v_{m_{peak}}$ power spectra.  The solid blue curve shows the $p$-value, neglecting the data points with $k<0.06 \ {\rm h \ Mpc^{-1}}$, and the dotted blue curve includes all of the low-$k$ points. The horizontal red line shows $p=0.05$. }}
     \label{newhalo}
   \end{figure}

\begin{figure}[htb] 
\centering
 \includegraphics[width=8.2cm]{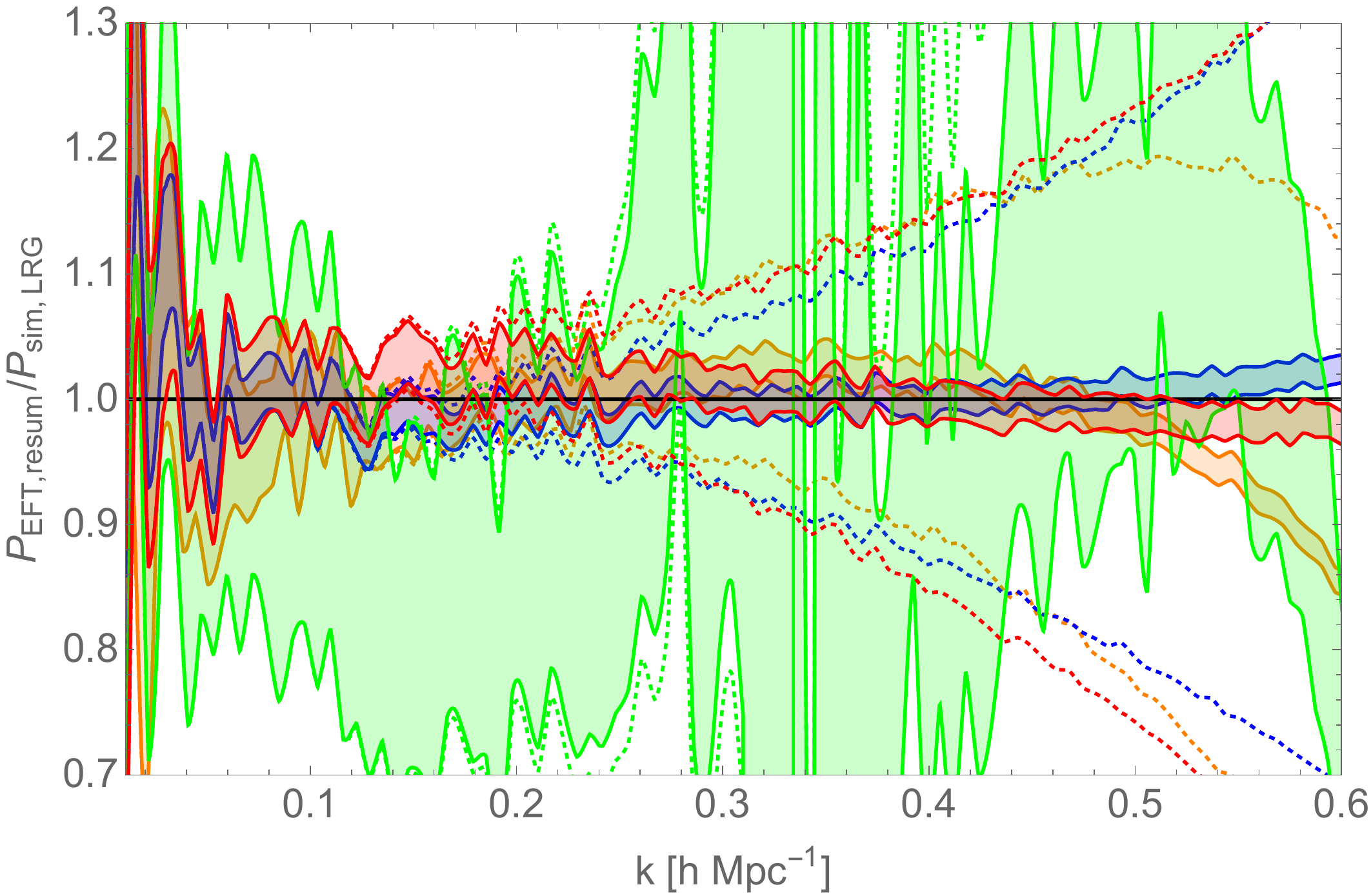}
 \includegraphics[width=8.2cm]{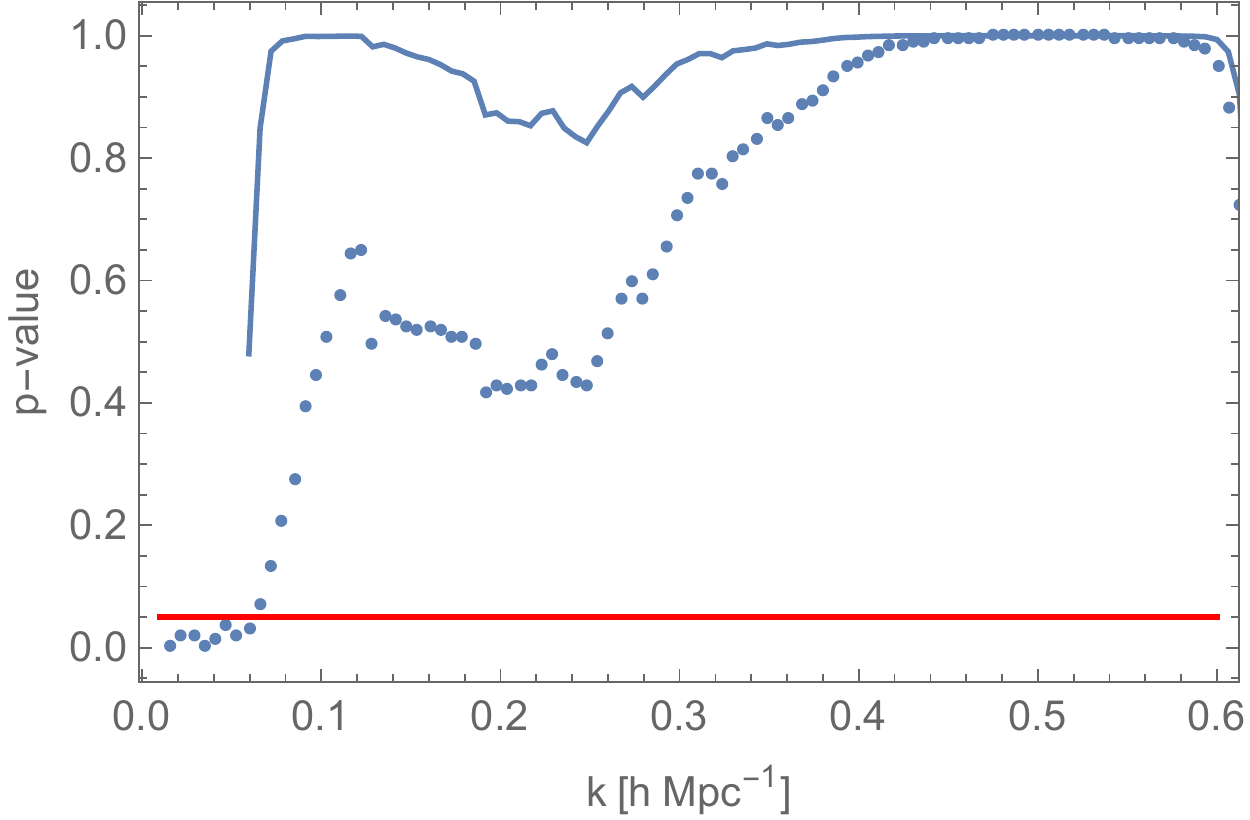}
  \caption{ {\it Left:} {Results of the fits, including theoretical error in quadrature, of the EFT power spectra at $z=0.67$ after IR-resummation to the power spectra of LRGs in the $v_{m_{peak}}$ sample~\cite{Jennings:2015lea}, which has a number density $\bar n=3.9 \cdot 10^{-4}(\hinvMpc)^3$, and dark matter extracted from simulations. The fits were performed  in the $k$-range $0.01 \ \unitsk$ to $0.54 \ \unitsk$ and resulted in the best-fit parameters $\{ b_1 = 1.94 \pm 0.03, b_2 = -0.08 \pm 9.06, b_3 = -0.81 \pm 3.89, b_4 = 1.7 \pm 7.2,   c_{\rm ct}^{(\delta_h)} = (9.9 \pm 10)\left( k_{\rm NL}  \ h^{-1} \ {\rm Mpc} \right)^{2}, \tilde{c}_{r,1} = (-39 \pm 4)\left( \km  \ h^{-1} \ {\rm Mpc} \right)^{2}, \tilde{c}_{r,2} = (9.4 \pm 1.4)\left( \km  \ h^{-1} \ {\rm Mpc} \right)^{2}, c_{\epsilon,1} = 4.5 \pm 63.0,   c_{\epsilon,2} = (18 \pm 6)\left( \km  \ h^{-1} \ {\rm Mpc} \right)^{2},  c_{\epsilon,3} = (35 \pm 9)\left( \km  \ h^{-1} \ {\rm Mpc} \right)^{2} \}$ for the LRGs and $c_s^2=(-0.49 \pm 0.01)\left( k_{\rm NL}  \ h^{-1} \ {\rm Mpc} \right)^{2}$ for the dark matter. $P_{\rm real}$ is plotted in red, $P_{l=0}$ in blue, $P_{l=2}$ in green, and $P_{\rm DM}$ in orange. The shaded region shows the 1$\sigma$ error on the simulation data, which includes the error on the spectra from simulations described in \cite{Feldman:1993ky} and a $1\%$ error added in quadrature to account for unknown systematics. The expected theoretical error is given by the dotted lines. {\it Right:} Plot of $p$-values calculated up to a given $k$ for the IR-resummed fit to the $v_{m_{peak}}$ power spectra.  The solid blue curve shows the $p$-value, neglecting the data points with $k<0.06 \ {\rm h \ Mpc^{-1}}$, and the dotted blue curve includes all of the low-$k$ points. The horizontal red line shows $p=0.05$. }}
     \label{newLRG}
   \end{figure}

\section{Details of Parameter Fits}\label{haloplots}
In this appendix we show the parameter plots used to determine the value of $k_{\rm fit}$ for the halos, as described in Section \ref{secfits}. We also include the correlation matrix of the parameters in Table~\ref{tb:corrmatrix1}.

\begin{figure}[htb!] 
\centering
\includegraphics[width=17cm]{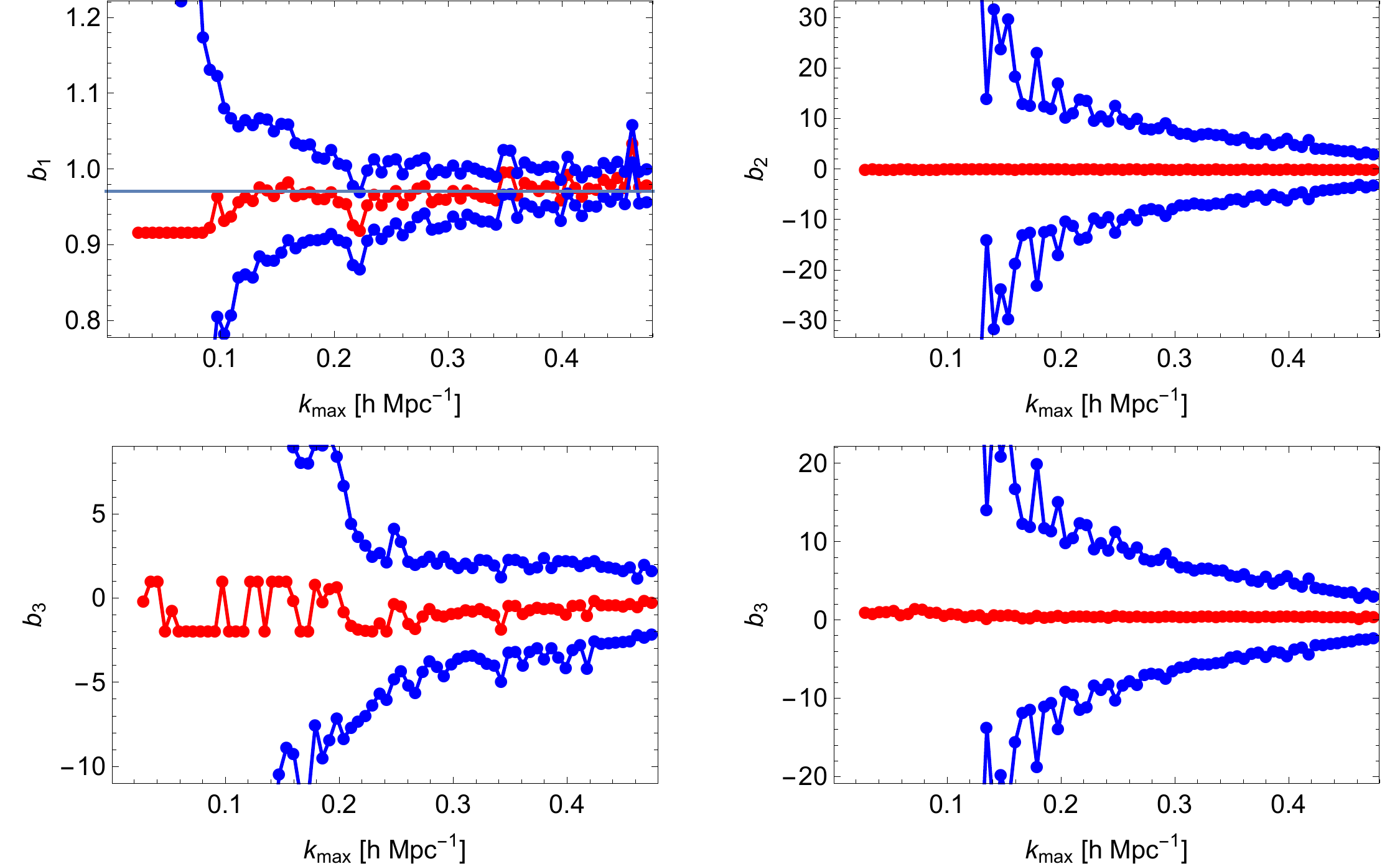} 
\caption{  Parameter plots with 2$\sigma$ errors for fits to halos up to a given $k_{\maxx}$, for bias parameters $b_1$, $b_2$, $b_3$, and $b_4$. The horizontal line shows the $k_{\maxx}$ at which $b_1$ becomes incompatible with the earlier values. }  \label{bplots1h}
\end{figure}
\begin{figure}[htb!] 
\centering
\includegraphics[width=17cm]{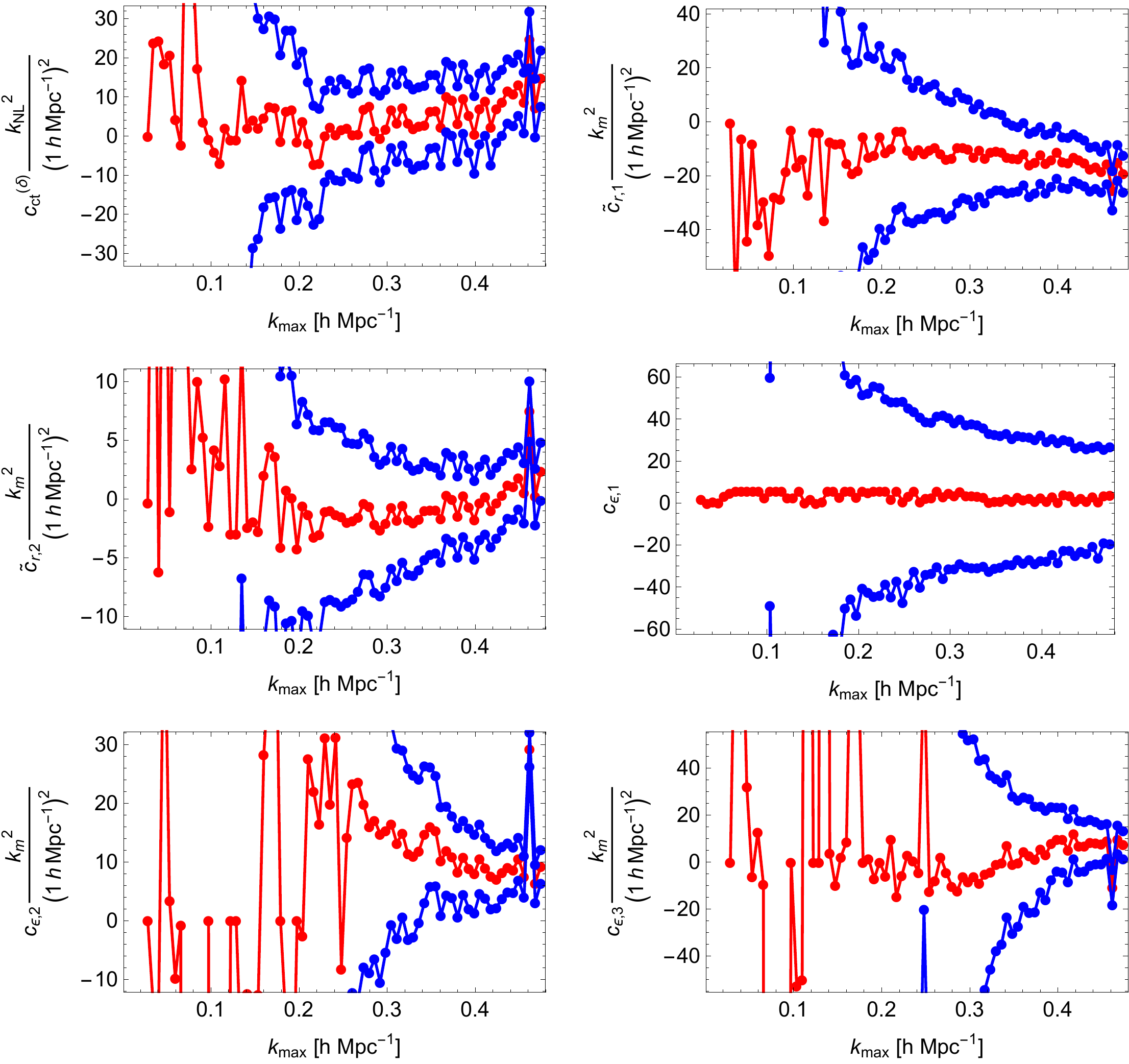} 
\caption{  Parameter plots with 2$\sigma$ errors for fits to halos up to a given $k_{\maxx}$, for counter-term parameters $c_{ct}^{(\delta)}$, $\tilde{c}_{r,1}$, $\tilde{c}_{r,2}$, $c_{\epsilon,1}$, $c_{\epsilon,2}$, and $c_{\epsilon,3}$. }  \label{bplots2h}
\end{figure}

\begin{table*}[h!]
\caption{{Correlation matrix for the bias coefficients of the halo sample.}}
\centering 
\setlength{\tabcolsep}{8pt}
\renewcommand{\arraystretch}{1.2}
\begin{tabular}{c|cccccccccc}
\hline\hline
\ & $b_1$ & $b_2$ & $b_3$ & $b_4$ & $c_{\rm ct}^{(\delta_h)}$ & $\tilde{c}_{r,1}$ & $\tilde{c}_{r,2}$ & $c_{\epsilon,1}$ & $c_{\epsilon,2}$ & $c_{\epsilon,3}$  \\
\hline
$b_1$ & 1. & 0.34 & -0.33 & -0.3 & -0.26 & -0.09 & -0.03 & -0.68 & -0.01 & 0.42 \\
\hline
$b_2$ &  0.34 & 1. & -0.98 & -1. & -0.84 & 0.87 & -0.77 & -0.87 & -0.77 & -0.45 \\
\hline
$b_3$ & -0.33 & -0.98 & 1. & 0.98 & 0.83 & -0.81 & 0.71 & 0.9 & 0.65 & 0.36 \\
\hline
$b_4$ &  -0.3 & -1. & 0.98 & 1. & 0.82 & -0.88 & 0.77 & 0.86 & 0.78 & 0.5 \\
\hline
$c_{\rm ct}^{(\delta_h)}$ & -0.26 & -0.84 & 0.83 & 0.82 & 1. & -0.81 & 0.91 & 0.65 & 0.72 & 0.14 \\
\hline
$\tilde{c}_{r,1}$ & -0.09 & 0.87 & -0.81 & -0.88 & -0.81 & 1. & -0.91 & -0.52 & -0.91 & -0.68 \\
\hline
$\tilde{c}_{r,2}$ & -0.03 & -0.77 & 0.71 & 0.77 & 0.91 & -0.91 & 1. & 0.45 & 0.86 & 0.39 \\
\hline
$c_{\epsilon,1}$ & -0.68 & -0.87 & 0.9 & 0.86 & 0.65 & -0.52 & 0.45 & 1. & 0.44 & 0.11 \\
\hline
$c_{\epsilon,2}$ & -0.01 & -0.77 & 0.65 & 0.78 & 0.72 & -0.91 & 0.86 & 0.44 & 1. & 0.63 \\
\hline
$c_{\epsilon,3}$ & 0.42 & -0.45 & 0.36 & 0.5 & 0.14 & -0.68 & 0.39 & 0.11 & 0.63 & 1. \\
\hline
\end{tabular}
\label{tb:corrmatrix1}
\end{table*}

\end{appendix}

\bibliography{references}

 \end{document}